\documentclass[11pt]{article}

\usepackage[margin=1in]{geometry}
\usepackage{graphicx}
\usepackage[textwidth=8em,textsize=small]{todonotes}
\usepackage{amsmath}
\usepackage{natbib}
\usepackage{dsfont}

\usepackage{amsthm}
\usepackage{amsfonts}
\usepackage{multirow}

\usepackage{amsmath,bbm,graphicx,amssymb,amsthm,amsbsy}
\PassOptionsToPackage{hyphens}{url}\usepackage{hyperref}
\hypersetup{
    colorlinks=true,
    linkcolor=blue,
    filecolor=magenta,      
    urlcolor=cyan,
    pdfpagemode=FullScreen,
    }
\usepackage{tabularx,booktabs}
\usepackage[utf8]{inputenc}
\usepackage{url}
\usepackage{enumitem}
\setlist[enumerate]{nosep}
\setlist[itemize]{nosep}

\usepackage[ruled]{algorithm2e}
\SetEndCharOfAlgoLine{}

\usepackage{setspace}
\makeatother

\newcommand\blfootnote[1]{%
  \begingroup
  \renewcommand\thefootnote{}\footnote{#1}%
  \addtocounter{footnote}{-1}%
  \endgroup
}

\begin{document}

\title{Variable importance measure for spatial machine learning models with application to air pollution exposure prediction}

\author{Si Cheng$^{1\dagger}$ \and 
Magali N. Blanco$^2$ \and
Lianne Sheppard$^{1,2}$ \and
Ali Shojaie$^{1\star}$ \and 
Adam Szpiro$^{1\star}$}
\date{
$^1$Department of Biostatistics, University of Washington\\
$^2$Department of Environmental \& Occupational Health Sciences, University of Washington
}

\maketitle
\blfootnote{$^\star$ indicates co-senior authors}
\blfootnote{$^\dagger$ contact: \href{mailto:si.cheng@aya.yale.edu}{si.cheng@aya.yale.edu}; Hans Rosling Center for Population Health, Box 351617, Seattle, WA 98195}
\vspace*{-5ex}
\begin{abstract}
Exposure assessment is fundamental to air pollution cohort studies. The objective is to predict air pollution exposures for study subjects at locations without data in order to optimize our ability to learn about health effects of air pollution. In addition to generating accurate predictions to minimize exposure measurement error, understanding the mechanism captured by the model is another crucial aspect that may not always be straightforward due to the complex nature of machine learning methods, as well as the lack of unifying notions of variable importance. This is further complicated in air pollution modeling by the presence of spatial correlation. We tackle these challenges in two datasets: sulfur (S) from regulatory United States national PM2.5 sub-species data and ultrafine particles (UFP) from a new Seattle-area traffic-related air pollution dataset. {Our key contribution is} 
a leave-one-out approach for variable importance that leads to interpretable and comparable measures for a broad class of models with separable mean and covariance components. {We illustrate our approach with several spatial machine learning models, and it clearly highlights the difference in model mechanisms, even for those producing similar predictions.} 
We leverage insights from this variable importance measure to assess the relative utilities of two exposure models for S and UFP that have similar out-of-sample prediction accuracies but appear to draw on different types of spatial information to make predictions. 


\end{abstract}

\section{Introduction}
\label{sec:intro}

Spatial prediction models are versatile tools that provide deeper understanding of social or natural mechanisms and guide decision making in practice. Examples include crime analysis in sociology \citep{chainey2008utility, zhao2017modeling, yi2018integrated}, nature disaster forecasting \citep{aggarwal1975spatial, arnaud2002influence, parker2017spatial, bui2018novel, karimzadeh2019spatial}, and exposure assessment in public health \citep{monn2001exposure, kibria2002bayesian, kim2009health, dias2018spatial, xu2022prediction}. 
The flexibility of machine learning (ML) models make them useful in prediction tasks with potentially complicated underlying mechanisms, but approaches to handling spatial structures in such models are relatively limited, compared to the abundance of ML methods, despite their practical importance. 

\citet{kanevski2009machine, li2011application, du2020advances} provided reviews and discussions on the application of ML models in spatial settings. 
Some approaches incorporate spatial information into the features that are used in vanilla ML models \citep[e.g.][]{kovacevic2009geological, cracknell2014geological, hengl2015mapping, hengl2018random}, which are straightforward to implement but do not provide explicit information on spatial heterogeneity and/or correlation; some combine ML methods and spatial smoothing into two-step models \citep[e.g.][]{bergen2013national, liu2018improve, chen2019kriging, blanco2021traffic}, which are flexible but may not partition the heterogeneity {attributable to the} mean and covariance components in an optimal way; and joint spatial-ML modeling \citep[e.g.][]{datta2016hierarchical, wai2020random, saha2021random, georganos2021geographical}, which are better-suited for spatial prediction, but may lead to more intensive computation and/or less clear theoretical properties.

Such considerations of pros and cons may guide the choice and interpretation of spatial ML models in practice. In addition to prediction accuracy and computational complexity, another crucial consideration is model interpretation, such as quantifying how much each predictor contributes to the predicted outcome {\citep[e.g.][]{xu2022prediction, masmoudi2020machine}}. This has been known as a challenge in ML literature partly due to the complexity of the models themselves; furthermore, the abundance of model classes and the wide variety of application disciplines {have given} rise to diverse and often incomparable measures of variable importance among different models, as noted by \citet{wei2015variable, greenwell2018simple, williamson2021nonparametric}. 
\citet{hooker2021unrestricted} summarized and described common challenges for different variable importance measures, most of which are intended for non-spatial models.
A common class of approaches is based on permuting the values of a covariate and assessing the change in prediction accuracy, as introduced by \citet{breiman2001random}. A related permutation-based approach was proposed by \citet{fisher2019all}, which considered averaging over all possible permutations. As noted by \citet{strobl2007bias, strobl2008conditional} and \citet{nicodemus2010behaviour}, results from these approaches may be questionable when features are correlated, as is often the case for spatial features which commonly come from GIS (geographic information system) data. \citet{friedman2001greedy, goldstein2015peeking} proposed using as a measure of variable importance the average predictions when all entries, or each entry, of a covariate take(s) a specific value. Another type of approach focuses on the change in predictive performance after re-fitting the model with the covariate of interest permuted, removed, or substituted \citep{mentch2016quantifying, candes2018panning, lei2018distribution}. 

Generalizing the approaches described above to spatial settings is non-trivial. 
First, {for approaches based on permuting or manually setting covariate values and evaluating on the model of interest, manipulating the covariates affects the predicted mean component of the given model, and consequently, the residuals are altered and may no longer be reasonably fitted by the previously-trained covariance model.}
In addition, the presence of spatial correlation will affect error estimates based on random sample splitting and in turn the validity of permutation-based approaches using out-of-bag observations \citep{russ2010spatial, meyer2019importance}. 
Furthermore, even when a valid variable importance measure can be presented as the change in prediction accuracy, interpretation from such approaches is limited since the exact quantitative contribution of each predictor on the outcome is still unclear.

Recognizing these challenges, we propose a leave-one-out approach based on quantile-level contrasts for variable importance in spatial ML models. This approach assesses the difference in predicted values for each data point when each predictor is fixed at different quantiles of its distribution. 
{Without refitting the whole model (which is often computationally intensive), the covariance component may not properly account for the change in the predictor values.}
Our proposed leave-one-out approach is flexible and efficient in that it examines each location individually for the spatial covariance component only, without requiring refitting of the computationally demanding mean model. It provides clear interpretation on how the change in each predictor, between different levels of its distribution, is associated with the difference in the outcome. This variable importance measure can be applied to a wide range of spatial machine learning models with separable mean and variance components, including multi-stage and joint models, and thus provides a standardized comparison between different modeling approaches.

The paper is organized as follows. {We introduce two air pollution datasets in Section~\ref{sec:data} with a focus on exposure assessment via spatial prediction. In particular, we argue that clearly different models could lead to highly similar patterns in predicted air pollution maps, and therefore additional information such as variable importance is crucial to comprehensive understanding and selection of models. To this end, we introduce a broad class of spatial ML models under a unifying framework in Section~\ref{sec:model}, and demonstrate how the models involved in Section~\ref{sec:data} fit into this framework as concrete examples. Under this modeling framework, we introduce our leave-one-out variable importance measure in Section~\ref{subsec:method}. Section~\ref{sec:analyses} illustrates the proposed approach on a synthetic dataset generated by a known mechanism, and presents variable importance analyses on the previously introduced air pollution studies.
}
We finish our discussion with some concluding remarks in Section~\ref{sec:disc}.
\section{Data Description}
\label{sec:data}

In this section, we introduce two air pollution datasets with air pollutant concentrations that are contained within a small and large geographic region, respectively. 
{Each dataset includes annual average concentrations of 5 and 4 air pollutants, along with measurements of 835 and 599 covariates, respectively. We build spatial prediction models for the concentration of each pollutant, and further seek to investigate the contribution of the covariates in each model.} 

\subsection{Seattle Mobile Monitoring Data} 
This study focused on characterizing annual average traffic-related air pollution (TRAP) levels in the greater Seattle area \citep{blanco2022characterization}, 
and leveraged a mobile monitoring (MM) campaign where a vehicle equipped with air monitors repeatedly collected two-minute samples at 309 stationary roadside sites. Approximately 29 measurements were collected from each site during all seasons, times of the week (weekdays, weekends), and most times of the day (5AM to 11PM) between March 2019 and March 2020. Prior work showed that this design provided unbiased annual average pollutant estimates \citep{blanco2022design}.

\begin{figure}
    \centering
    \includegraphics[width=14cm]{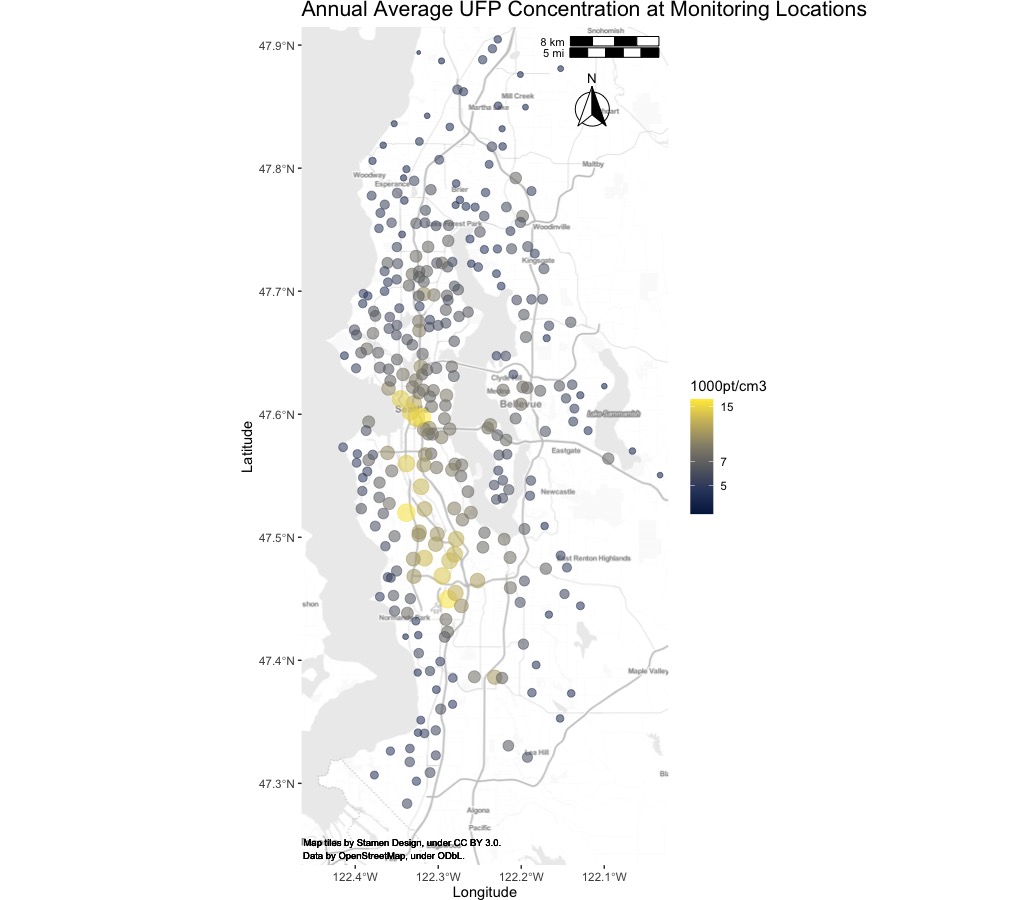}
    \caption{ (Estimated) annual average concentration of UFP from Seattle mobile monitoring data. The color and size of dots both reflect the magnitude of concentration. If not noted otherwise, all maps in this paper were made using the \texttt{ggmap} R package \citep{Kahle2013ggmap}. Map tiles by Stamen Design, under CC BY 3.0; data by OpenStreetMap, under ODbL.}
    \label{fig:sea_annavg}
\end{figure}

Measured pollutants included ultrafine particles (UFP), black carbon (BC), nitrogen dioxide (NO\textsubscript{2}), carbon dioxide (CO\textsubscript{2}) and fine particulate matter (PM\textsubscript{2.5}).
Median 2-minute visit concentrations were trimmed at the site level such that concentrations below the 5\textsuperscript{th} and above the 95\textsuperscript{th} quantile for a given site site were removed. This was done to reduce the influence of large outlier concentrations prior to calculating annual average site concentrations.
Figure~\ref{fig:sea_annavg} visualizes the resulting annual average UFP concentrations. Annual average concentration for other pollutants are presented in Appendix~\ref{app:annavg}. 
We focus our discussion on UFP in the main text, although the results for other pollutants {lead to similar conclusions and} are included in Appendix~\ref{app:annavg}. 
Pollutant concentrations were log-transformed prior to model-fitting, and transformed back to the original scale to assess the $R^2$ (and variable importance in Section~\ref{sec:analyses}).

\subsection{National PM\textsubscript{2.5} Sub-Species Monitoring Data}
This dataset consists of measurements of four PM\textsubscript{2.5} sub-species, elemental carbon (EC), organic carbon (OC), silicon (Si), and sulfur (S) across the United States, and was collected during 2009 -- 2010 by two  U.S. Environmental Protection Agency networks: the Interagency Monitoring for Protected Visual Environments (IMPROVE) and Chemical Speciation Network (CSN) \citep{sacks2009integrated}. 
Following the approaches in \citet{bergen2013national} and \citet{wai2020random}, we only included in our analyses measurements from the monitors that had at least 10 data points per quarter and a maximum of 45 days between measurements. {We calculated annual average concentrations of S and Si from from 323 IMPROVE and CSN monitors over 01/01/2009 -- 12/31/2009. For EC and OC, we averaged measurements from 204 IMPROVE and a subset of CSN monitors over 01/01/2009 -- 12/31/2009 and from the remaining 51 CSN monitors over 05/01/2009 -- 04/30/2010. The later averaging period was used for some of the CSN monitors due to a change in the measurement protocol. See \citet{bergen2013national} for additional details.}
We focus our discussion on the modeling of S, for which the annual average concentration is plotted in Figure~\ref{fig:national_annavg}. Annual averages and results for EC, OC and Si are presented in Appendix~\ref{app:annavg}.
For consistency and comparability to previous analyses of the same data \citep{bergen2013national, wai2020random}, we square-root transformed the annual averages before modeling, and then transformed them back to the natural scale before presenting results.

\begin{figure}
    \centering
    \includegraphics[width=15cm]{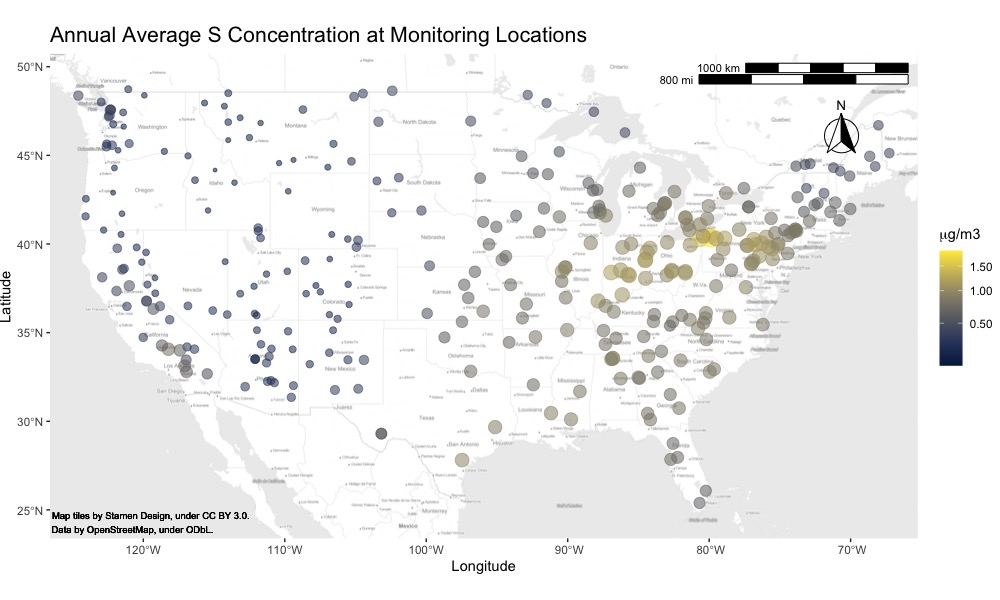}
    \caption{ (Estimated) annual average concentration of S from national PM\textsubscript{2.5} monitoring data. The color and size of dots both reflect the magnitude of concentration.}
    \label{fig:national_annavg}
\end{figure}

\subsection{Geographic Covariates}
Information on 835 geographical covariates 
for the Seattle TRAP data, and 599 covariates for the national PM\textsubscript{2.5} data was available from the MESA Air (Multi-Ethnic
Study of Atherosclerosis and Air Pollution) Database \citep{mesa2019} at all monitor locations within each dataset. Table~\ref{tab:geocov} presents the types, details and sources of geographical information on these covariates.

We pre-processed the covariates as described in \citet{keller2015unified}. Specifically, geocovariates that lacked variability (less than 20\% of the data were different from the most common value) or had too many outliers ($>2\%$ of the sample size) were excluded; proportion land use variables were excluded if the maximum proportion observed in the dataset was less than 10\%.  This led to a total of 183 and over 480 (482 covariates for EC and OC, and 489 covariates for S and Si, where the difference is because of slightly different monitoring locations for different pollutants) geocovariates for the analysis of the Seattle TRAP and national PM\textsubscript{2.5} datasets, respectively.

\begin{table}[]
\small
\begin{tabular}{@{}p{0.2\linewidth}p{0.2\linewidth}p{0.2\linewidth}p{0.35\linewidth}@{}}
\toprule
Covariate Category & Data Source                                                                                              & Buffer Sizes                                                                       & Notes                                                                                                                                                                \\ \midrule
Airports           & National Emissions Inventory Database                                                                                             & --                                                                                 & Distances to airports and large airports                                                                                                                             \\
Coastline          & TeleAtlas                                                                                                & --                                                                                 & Distance to coastline                                                                                                                                                \\
Railroads          & TeleAtlas                                                                                                & --                                                                                 & Distance to railroads                                                                                                                                                \\
Railyards          & TeleAtlas                                                                                                & --                                                                                 & Distance to railyards                                                                                                                                                \\
Roads              & TeleAtlas                                                                                                & --                                                                                 & Distances to A1, A2 and A3 roads                                                                                                                                     \\
                   &                                                                                                          & 50m -- 5km                      & Lengths of A1, A2 and A3 roads within a buffer                                                                                                                                          \\
Intersections\textsuperscript{*}      & TeleAtlas                                                                                                & --                                                                                 & Distances to A1/A2/A3 intersections                                                                                                                                  \\
                   &                                                                                                          & 500m, 1km, 3km                                                                     & Number of A1/A2/A3 intersections within a buffer                                                                                                                                   \\
Population   & US Census Bureau                                                                                         & 500m -- 15km                                        & Population within a buffer                                                                                                                                           \\
Land use           & MRLC 2006 National Landcover Dataset \& USGS historical source & 50m -- 15km          & Land use (e.g. commercial, residential, urban, cropland, mixed forest, streams, beaches) within a buffer                                                                                              \\
                   & USGS historical source                                                                                   & --                                                                                 & Distance to commercial and services land use                                                                                                                                      \\
Ports              & National Geospatial Intelligence Agency                                                                  & --                                                                                 &  Distances to small, medium,
large ports     \\
Emission Sources   & National Emissions
Inventory Database                                                                                             & 3km, 15km, 30km                                                                    & Sum of major emissions NOx, SO\textsubscript{2}, PM\textsubscript{2.5}, CO and PM\textsubscript{10} within a buffer                                                                                                 \\
Truck routes\textsuperscript{*}       & Bureau of Transportation Statistics                                                                      & --                                                                                 & Distance to truck routes                                                                                                                                             \\
                   &                                                                                                          & 50m -- 15km          & Length of truck routes within a buffer                                                                                                                               \\
Impervious Surface\textsuperscript{*} & National Landcover Dataset                                                                               & 50m -- 5km                             &    Percentage of an area covered with an impervious surface (e.g. pavement, concrete)
within a buffer  \\
Elevation\textsuperscript{*}          & National Elevation Dataset                                                                               & --                                                                                 & Elevation above sea level                                                                                                                                            \\
                   &                                                                                                          & 1km, 5km                                                                           & Relative elevation: counts of points within a buffer that is less/more than 20m/50m uphill/downhill of the location                                                  \\
Normalized difference vegetation index (NDVI)          & University of Maryland                                                                                   & \begin{tabular}[c]{@{}l@{}}250m -- 10km\end{tabular} & Measures the level of vegetation in a monitor’s vicinity; summarized at: the 25th/50th/75th percentiles annually; median of summer (Apr to Sept) and winter (Jan to Mar and Oct to Dec) \\ \bottomrule
\end{tabular}
\caption{Summary of available geographical information. Distances to spatial features were truncated at 25km in the Seattle TRAP data, and at 10km in the national data. All these geocovariates were available for the Seattle mobile monitoring locations, while those marked with asterisks (*) were not available at the IMPROVE and CSN monitoring locations, and thus not included in the national study.}
\label{tab:geocov}
\end{table}

\subsection{Predicting Pollutant Concentration}
\label{subsec:data-pred}
As a motivation for our investigation of variable importance, we first discuss the prediction of pollutant concentrations on the Seattle and national air pollution datasets. {We briefly describe two main prediction approaches below, and then introduce them rigorously under a unifying framework in Section~\ref{sec:model}.}  We conducted 10-fold cross-validation and characterized the performance of these models via $R^2$.
\begin{itemize}
    \item UK-PLS: a two-step procedure that first extracts the top partial least squares (PLS) \citep{wold1984collinearity, sampson2011pragmatic} components from the covariates (where the number of components is determined by cross-validation within the training set, with the goal of maximizing prediction $R^2$), and then fits a universal kriging model via maximum likelihood with the selected components as covariates and an exponential covariance structure; 
    \item SpatRF (PL): a spatial random forest algorithm proposed by \citet{wai2020random}, where the tree-building algorithm selects each split of the tree adjusted for spatial correlation via thin plate regression splines (TPRS). {We adopted the pseudo-likelihood (PL) optimization approach introduced in \citet{wai2020random}}, and selected the hyperparameters by grid search via cross-validation. 
\end{itemize}

\begin{figure}[!ht]
    \centering
    \includegraphics[width=17cm]{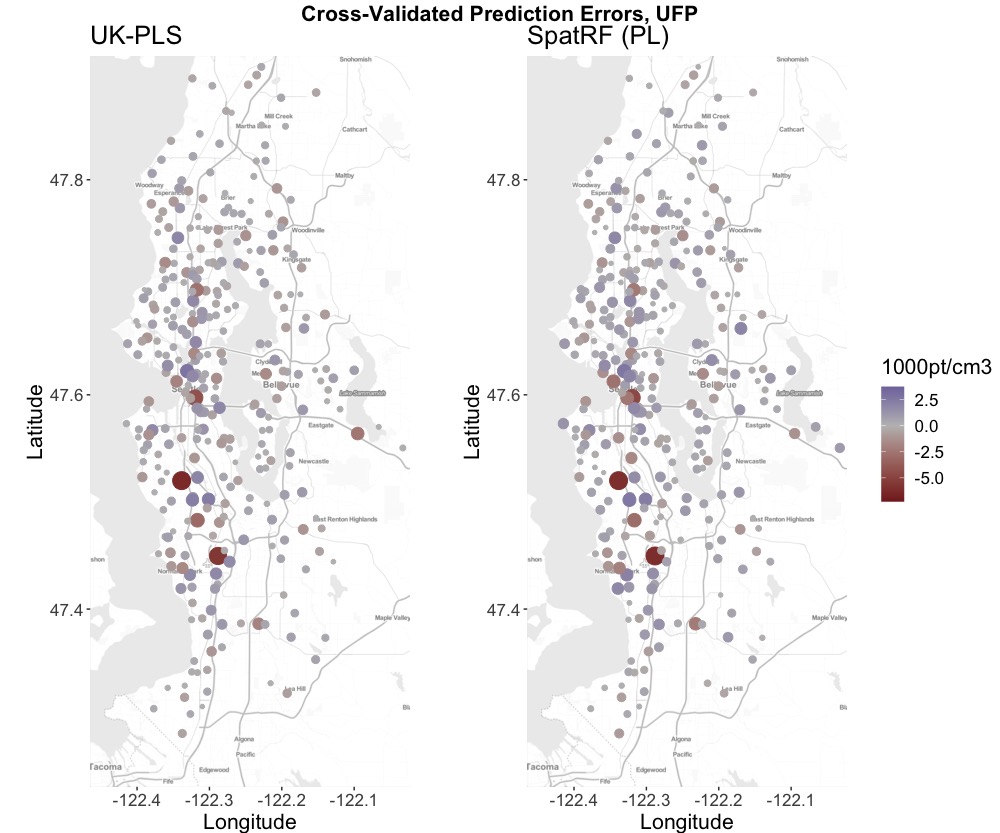}
    \caption{Cross-validated prediction errors of UFP for UK-PLS and spatial RF (PL) at each monitoring location of the Seattle study. The shade of color and size of dots both reflect the magnitude of errors.}
    \label{fig:sea_cverr}
\end{figure}

\begin{figure}[!ht]
    \centering
    \includegraphics[width=16.5cm]{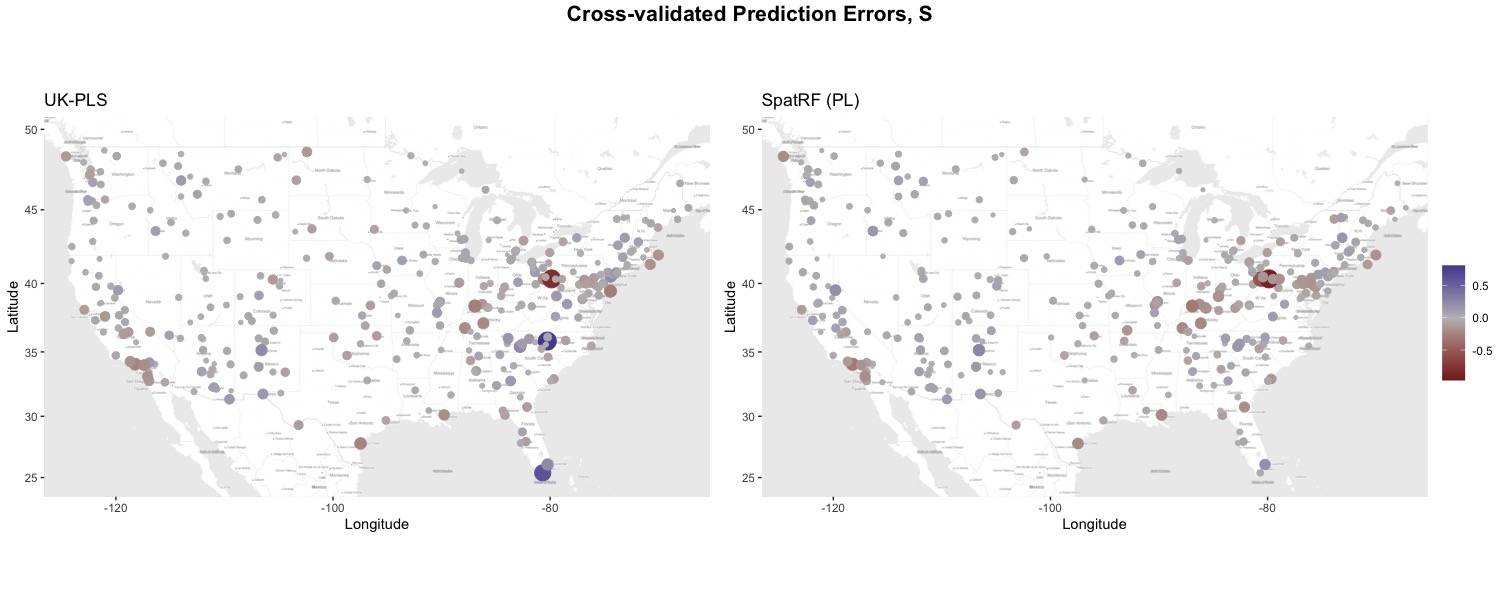}
    \caption{Cross-validated prediction errors of S for UK-PLS and spatial RF (PL) at each monitoring location of the national study. The shade of color and size of dots both reflect the magnitude of errors.}
    \label{fig:national_cverr}
\end{figure}

For the prediction of UFP concentration with the Seattle data, UK-PLS and spatial RF (PL) achieve cross-validated $R^2$'s of 0.81 and 0.78, respectively. 
Figure~\ref{fig:sea_cverr} displays the cross-validated prediction errors of UK-PLS and spatial RF for UFP at all monitoring locations in this study. 
For the national data where we predicted Sulfur concentration, 
the $R^2$ of UK-PLS and spatial RF (PL) are 0.89 and 0.90, respectively; the cross-validated prediction errors are displayed in Figure~\ref{fig:national_cverr}.

\begin{figure}[!ht]
    \centering
    \includegraphics[width=16.5cm]{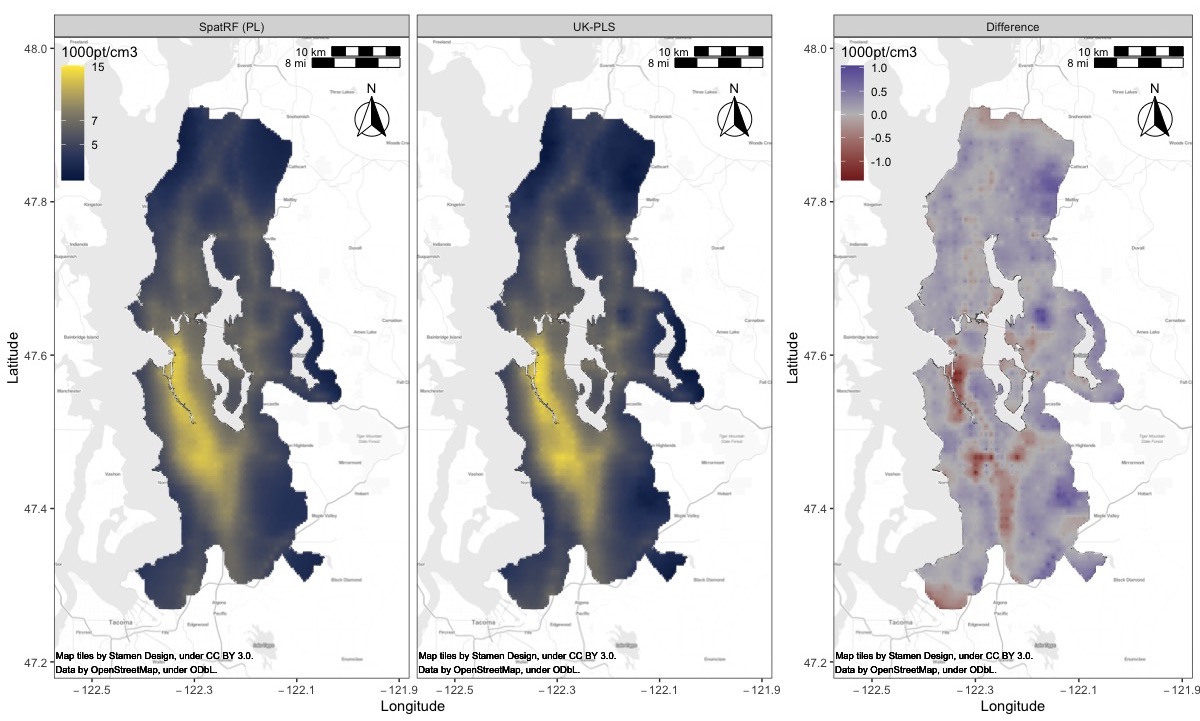}
    \caption{Predicted UFP concentration surfaces based on predictions at gridded locations via UK-PLS and Spatial RF (PL) in the Seattle TRAP study region, along with a difference map between them (UK-PLS being the subtrahend) }
    \label{fig:sea_grid}
\end{figure}

For the Seattle data, we observe highly similar spatial patterns in the distribution of prediction errors across the monitoring locations produced by UK-PLS and spatial RF, despite their clearly different nature: UK-PLS captures a linear trend in the mean model while spatial RF allows for non-linear effects; UK-PLS is a two-step procedure with explicit dimension reduction followed by spatial smoothing, while spatial RF conducts implicit degree-of-freedom control and jointly accounts for the mean and covariance components. 
On the other hand, the gridded prediction maps over the Seattle TRAP study region is shown in Figure~\ref{fig:sea_grid}, which is based on the evaluation of each model at an additional set of 2815 gridded locations (with higher resolution) within the same study region. The difference map on the third panel reveals that predictions made by UK-PLS and Spatial RF which are highly similar at the mobile monitoring locations, when extrapolated to a higher resolution, could still exhibit different spatial patterns.
On the predicted concentrations of S, we see that while both models achieved similar accuracy and produced similar predictions for locations in mid- and western US, their different behaviors in eastern US were reflected by the larger (positive) prediction errors of UK-PLS at a few locations. 

All these observations indicate that different spatial prediction models may appear to be highly similar when restricted to certain areas, while the true underlying difference between them may not be observed merely based on their predictions, if evaluations at an additional set of locations (e.g. the gridded locations in addition to the mobile monitoring locations in the Seattle data, or eastern US comparing to mid- and western US in the national data) were unavailable. 
It would therefore be desirable and also necessary to understand and compare different models by investigating the mechanisms that they capture, beyond just their prediction performance on the training data.
Developing a universal and easily interpretable variable importance measure for a diverse class of prediction methods is a key step to facilitate this, {and further to aid the selection and interpretation of models.}

\section{Variable Importance Measure for Spatial ML Models}
\label{sec:method}

\subsection{Spatial Prediction: Setup}
\label{sec:model}
Before introducing the proposed variable importance measure, we describe a broad class of spatial prediction models to which such measure is applicable.
Consider a class of models where an outcome $Y(s)$, indexed by location $s\in \Omega$, is modeled via an additive mean surface taking the form of

\begin{equation}
    g(\mu(s)) = \sum_{k=1}^K [f_k\left(X(s)\right) + \nu_k(s)]
    \label{equ:meanmod}
\end{equation}
where $g(\cdot)$ is a link function, $X(s)$ represents the covariates, each $\nu_k(s)$ represents the correlated error term, and each $f_k(\cdot)$ is an unknown function within some function class $\mathcal F$. The indexing $k$ allows for application to ensemble learning methods.
As an example, when $g$ is the logarithm link function and $\nu(s)$ is a correlated Gaussian process, $\mu(s)$ is the underlying intensity of a doubly-stochastic Poisson process, also known as the Cox process \citep{cox1955some, serfozo1972conditional, bremaud1981point}; when $g$ is the identity link and $\nu(s)$ is a correlated Gaussian process, $\mu(s)$ models the surface of a continuous outcome, e.g. a universal kriging model if $f$ is linear.

A spatial ML model often learns about each $f_k$ under some assumed restrictions on $\mathcal F$, and $\nu_k(s)$ under parametric or structural assumptions. Once a model (\ref{equ:meanmod}) is fitted at a set of training sites $s_{\text{trn}}$, predictions at the test sites $ s_{\text{tst}}$ can be made via
\[
    \hat Y( s_{\text{tst}}) = g^{-1} \left(\sum_{k=1}^K \left\{ \hat f_{k}(X( s_{\text{tst}})) + \mathbb E_{\hat\nu} \left[\nu_k (s_{\text{tst}}) \mid Y(s_{\text{trn}}), \hat f_k(s_{\text{trn}}) \right] \right\}\right),
\]
where the conditional expectation $\mathbb E_{\hat\nu} \left[ \nu_k (s_{\text{tst}}) \mid Y(s_{\text{trn}}), \hat f_k(s_{\text{trn}}) \right]$ represents the smoothing of residuals {via the fitted covariance model $\hat\nu_k(s)$} at the training sites. When $g$ is identity and each $\nu_k$ is assumed to be a correlated Gaussian random field with covariance $\Sigma(\theta)$, for example, the model predicts
\begin{equation*}
    \hat\nu_k (s_{\text{tst}}) := \mathbb E_{\hat\nu} \left[ \nu_k (s_{\text{tst}}) \mid Y(s_{\text{trn}}), \hat f_k(s_{\text{trn}}) \right] = \Sigma(\hat\theta)_{\text{tst,trn}} \left[\Sigma(\hat\theta)_{\text{trn,trn}} \right]^{-1} \left(Y(s_{\text{trn}}) - \hat f_k(s_{\text{trn}})
    \right).
\end{equation*}

As an illustration and for concreteness of our following discussions, we revisit two the models introduced in Section~\ref{subsec:data-pred} for an observed continuous outcome $Y_{n\times 1}$, with identity $g(\cdot)$ and potentially high-dimensional covariates $X_{n\times p}$, and describe how they fit into this framework.
One model is UK-PLS, which {has $k = 1$ and} first conducts PLS and extracts the first $l$ ($1\le l \le p$) components of $X$ by finding a decomposition of $X$
\begin{align*}
    T_{n\times l} := X_{n\times p} H_{p\times l}
\end{align*}
such that the covariance between $T$ and $Y$ is maximized. The number $l$ of components to use can be selected via cross-validation. In the second step, a universal kriging model
\begin{align*}
    Y_{n\times 1} &= T_{n\times l}\beta_{l\times 1} + \nu_{n\times 1} \\
    \nu_{n\times 1} &\sim \text{Normal}(0, \Sigma(\theta))
\end{align*}
where $\theta$ are covariance parameters (e.g. the nugget, partial sill and range, see \citet{cressie2015statistics}) which can be solved jointly with $\beta$ via maximum likelihood. With $\hat H_{p\times l}, \hat\beta, \hat\theta$ estimated from the model, for $m$ new (test) locations with covariate values $X^*_{m\times p}$, the outcome $Y^*$ can be predicted as
\begin{align*}
    \hat Y^* = X^*\hat H\hat\beta + \mathbb E_{\theta}[\nu^*\mid \nu]
    = X^*\hat H\hat\beta + \tilde\Sigma_{12}(\hat\theta) \tilde\Sigma_{22}^{-1}(\hat\theta) 
    \left( Y - T\hat\beta \right)
\end{align*}
where $\tilde\Sigma_{(m+n)\times (m+n)}(\hat\theta)$ is the covariance matrix induced by the distances between all training and test locations, partitioned as
\begin{equation}
    \tilde\Sigma(\hat\theta) = \begin{bmatrix}
        \tilde\Sigma_{11}^{(m\times m)} & \tilde\Sigma_{12}^{(m\times n)}\\
        \tilde\Sigma_{21}^{(n\times m)} & \tilde\Sigma_{22}^{(n\times n)}
    \end{bmatrix}
    \label{equ:Sig-partition}
\end{equation}
based on the training and test indices. 

The second example is the spatial random forest algorithm proposed by \citet{wai2020random} solved via pseudo-likelihood (SpatRF-PL). It {is an ensemble model (i.e. $k > 1$) which} specifies 
\begin{equation}
    \hat\mu(s) := \sum_{k=1}^K \hat\mu_k(s) := \sum_{k=1}^K \left[ \hat f_k(X(s)) + \hat\nu_k(s) \right]
    \label{equ:spatrf}
\end{equation}
where each $f_k$ is a regression tree, and each $\nu_k$ could be modeled via common spatial smoothing methods such as kriging or regression splines \citep{friedman1991multivariate, wood2003thin}. 
With a kriging model, for each $k$, a spatially adjusted tree can be built by solving the optimization problem resulting from profile likelihood, assuming normally distributed spatial error terms:
\begin{align}
    \notag \arg\max_{\theta_k} & \left[-\frac{1}{2}\log\lvert\Sigma(\theta_k)\rvert
    -\frac{1}{2}\left(Y - \hat f_k(X\mid \Sigma(\theta_k)) \right)^\top \Sigma^{-1}(\theta_k)\left(Y - \hat f_k(X\mid \Sigma(\theta_k)) \right)
    \right]\\
    & \text{s.t. } \hat f_k(X\mid\Sigma(\theta_k)) = \arg\min_{f_k(X\mid\Sigma(\theta_k))}
    \left(Y -  f_k(X\mid \Sigma(\theta_k)) \right)^\top \Sigma^{-1}(\theta_k)\left(Y -  f_k(X\mid \Sigma(\theta_k)) \right).
    \label{equ:spat-tree}
\end{align}

And likewise, predictions at test locations can be made via
\[
    \hat Y^* = \sum_{k=1}^K \left[
        \hat f_k(X^*) + \mathbb E_{\hat\theta_k} \left(\nu^*\mid \nu\right)
    \right]
    = \sum_{k=1}^K \left[
        \hat f_k(X^*) + \tilde\Sigma_{12}(\hat\theta_k) \tilde\Sigma_{22}^{-1}(\hat\theta_k) 
    \left( Y - \hat f_k(X) \right)
    \right]
\]
with $\tilde\Sigma$ defined identically as (\ref{equ:Sig-partition}).

\subsection{Leave-One-Out Evaluation of Quantile-Level Contrasts}
\label{subsec:method}

We now introduce a variable importance measure that is applicable to additive models taking the form of (\ref{equ:meanmod}) as described in Section~\ref{sec:model}. This leave-one-out approach is based on the change in predictions across different user-specified quantiles $q_1, \ldots, q_m$ for each covariate, evaluating at each location $s_1, \ldots, s_n$ one at a time.
{Recall that prediction at the test locations ${s}_{\text{tst}}$ relies on evaluation of the trained covariance model $\hat\nu(s)$ via $\hat\nu(s_{\text{tst}})=\mathbb E[\nu(s_{\text{tst}}) \mid \nu(s_{\text{trn}}) = Y(s_{\text{trn}}) - \hat f(s_{\text{trn}})]$. This implies that when we permute or fix the values of covariates $X_{\text{tst}}$ as in many common variable importance analyses, the evaluation of the covariance model $\hat\nu(s)$ is also implicitly altered, and furthermore, the distribution of residuals $\nu(s_{\text{tst}})$ at the test locations may no longer be well-fitted by $\hat\nu(s)$.}
Therefore, the key idea of the proposed approach is to reuse the trained mean model across all locations, but re-fit the covariance model in a leave-one-out manner. 
We write $\hat F_{X_j}(x_j), j = 1, \ldots, p$ as the empirical cumulative distribution function (CDF) of the $j$th covariate, and $\boldsymbol s_{-i}$ as the set of all locations except the $i$th one. 

Suppose we have trained a model
\[
    g\left(\hat\mu(s)\right) = \sum_{i=1}^K \left[\hat f_k(X(s)) + \hat\nu_k(s) \right]
\]
from observations $\{(X(s_i), Y(s_i))\}_{i=1}^n$. Then for the $j$th covariate, at the $l$th quantile $q_l$ of interest and within the $k$th sub-model $\hat f_k$, we replace each $X_j(s_i)$ with the sample $q_l$-quantile and calculate the predicted mean as
\[
    \hat\zeta_{k}^{j,l}(s_{i}) := \hat f_k\left(X_1(s_{i}), \ldots, \hat F_{X_j}^{-1}(q_l), \ldots, X_p(s_{i})\right)
\]
for location $i$. In plain words, this is the {new} predicted mean at $s_i$ with the $j$th covariate replaced by its $q_l$-th quantile across $s_1, \ldots, s_n$. 
Next, we re-fit the $k$th error component with the {new} predicted means $\hat\zeta_{k}^{j,l}({\boldsymbol s_{-i}})$ along with observations $\left(X({\boldsymbol s_{-i}}), Y({\boldsymbol s_{-i}})\right)$, leaving out the $i$th site. Denoting the resulting model as $\hat\nu_{(-i),k}^{j,l}(s)$, the leave-one-out approach yields the linear predictor
\begin{equation}
    \hat\eta_{k}^{j,l}(s_i) := \hat\zeta_{k}^{j,l}(s_i) + \mathbb E_{\hat\nu_{(-i),k}^{j,l}}\left[\nu_{(-i), k}^{j, l}(s_i)\mid Y(\boldsymbol s_{-i}), \hat f_k\left(X(\boldsymbol s_{-i})\right)\right]
    \label{equ:linear-pred}
\end{equation}
for location $i$, which is what the model would predict if the $j$th covariate of all data points were replaced by the $q_l$-quantile of its distribution, and if the error component were fitted without the $i$th data point, while keeping everything else intact. {Re-fitting leads to updated covariance model(s) that account for the implicit change in the error distribution caused by manipulating the covariates.}

Re-doing this for all $i$, we obtain a sequence of linear predictors of the form (\ref{equ:linear-pred}). Aggregating across each sub-model (each $k$) and location (each $i$) finally leads to the averaged leave-one-out predictions at the $q_l$-th quantile for covariate $j$:
\begin{equation}
    \bar\mu_{j,l} := g^{-1}\left(\frac{1}{n} \sum_{i=1}^n \sum_{k=1}^K \hat\eta_{k}^{j,l}(s_i) \right).
\end{equation}

For each $j$, the trajectory $\bar\mu_{j,1}, \ldots, \bar\mu_{j,m}$ reflects how the predictions, on average, vary across different quantiles of covariate $X_j$, which serves as an intuitive measure of the contribution of this covariate on the predicted outcome. This procedure could easily be parallelized to facilitate computation. Algorithm~\ref{algo:varimp} presents the described procedure in detail. 

\RestyleAlgo{boxruled}
\SetAlgoNoLine
\begin{algorithm}[!ht]
  \caption{Leave-one-out variable importance\label{algo:varimp}}
  Input: data $(X_{n\times p}, Y_{n\times 1})$, quantile levels $q_1, \ldots, q_m \in [0, 1]$ of interest, and a trained model 
  \[
    g\left(\hat\mu(s)\right) = \sum_{k=1}^K \left[\hat f_k(X(s)) + \hat\nu_k(s) \right]
  \]\;
            \For{$j = 1, \ldots, p$}{
                \For{$l = 1, \ldots, m$}{
                \For{$k = 1, \ldots, K$}{
                    \For{$i' = 1, \ldots, n$}{
                {Replace the $i'$th observation of the $j$th covariate, $X_{i'j}$, with the $q_l$-th sample quantile $\hat F_{X_j}^{-1}(q_l)$}:
                    \[
                        \tilde X_{i'\cdot} := \left(X_1(s_{i'}), \ldots, \hat F_{X_j}^{-1}(q_l), \ldots, X_p(s_{i'})\right)
                    \]

                {Calculate the new predicted mean as $\hat\zeta_k^{j,l}(s_i'):=\hat f_k(\tilde X_{i'\cdot})$}\;}

                \For{$i = 1, \ldots, n$}{ 
                Re-fit a covariance model on $(X_{-i,\cdot}, Y_{-i})$ with the updated residuals $Y(\boldsymbol s_{-i}) - \hat\zeta_k^{j,l}(\boldsymbol s_{-i})$, denoted as $\hat\nu_{(-i),k}^{j,l}(s)$\;
        
                    
                    Evaluate the covariance term for location $i$ from the re-fitted covariance model $\hat\nu_{k}^{j,l}(s_i) := \mathbb E_{\hat\nu_{(-i),k}^{j,l}} \left[\nu_{(-i), k}^{j,l}(s_i)\mid Y(\boldsymbol s_{-i}), \hat\zeta_k^{j,l}\left(\boldsymbol s_{-i}\right)\right]$\;
                    
                    Calculate the linear predictor for location $i$ as $\hat\eta_{k}^{j,l}(s_i) := \hat\zeta_{k}^{j,l}(s_i) + \hat\nu_{k}^{j,l}(s_i)$\;
                    
                }
            }
            Calculate the averaged leave-one-out predictions $\bar\mu_{j,l} := g^{-1}\left(\frac{1}{n} \sum_{i=1}^n \sum_{k=1}^K \hat\eta_{k}^{j,l}(s_i) \right)$\;
        }
    }
    \KwResult{Output averaged leave-one-out predictions $\bar\mu_{j,l}$ for each $j = 1, \ldots, p$, and $l = 1, \ldots, m$. }
\end{algorithm}
\section{Variable Importance Analyses}
\label{sec:analyses}

\subsection{Illustration with Synthetic Data}
\label{subsec:syn-data}

We first illustrate how the proposed variable importance measure performs in comparison to the true mechanism, by presenting a variable importance analysis with synthetic data generated with the same covariates as the national study. We generated a continuous outcome with five active predictors
\begin{itemize}
    \item Distance to A1 road
    \item Population density
    \item Annual median NDVI, buffer size 1km
    \item Land use: mixed urban, buffer size 15km
    \item Land use: residential, buffer size 15km
\end{itemize}
which were all scaled and centered, except population density and mixed urban land use which were first scaled and then shifted to be non-negative. The mean model was given by
\begin{align*}
    \mathbb E(Y\mid X) = & -0.5\times \text{distance to A1 road} + 0.2\times \text{population density}^2 - 1\times \text{annual median NDVI}\\
    & + 0.5\times\sqrt{\text{mixed urban land use}} + 0.5\times \text{residential land use}\\
    & -0.25\times \text{distance to A1 road}\times\text{annual median NDVI}. 
\end{align*}

The correlated error term was given by an exponential model with scale and range parameters equal to 4 and 2.5 respectively, where the unit of distance was 1000km. The uncorrelated errors were generated from a standard Gaussian distribution. With this setup, the variances of each component (mean, partial sill and nugget) in the outcome were 2.21, 1.07 and 1.02 respectively, so that we roughly have a 2:1:1 variance decomposition of the outcome. {The distribution of this synthetic outcome and variability coming from each component are visualized in Figure~\ref{fig:y_decomp} in Appendix~\ref{app:varimp}.}

{We trained both UK-PLS and spatial RF to predict this synthetic outcome with all covariates. Although there is an interaction term in the true mechanism, only main effects were included when training each model. UK-PLS and spatial RF achieve cross-validated $R^2$ 0.62 and 0.72 respectively.}

\begin{figure}
    \centering
    \includegraphics[width=16.5cm]{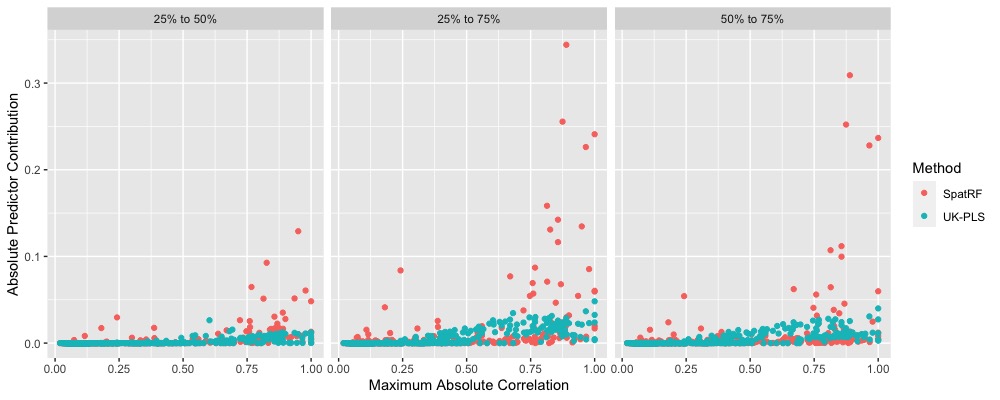}
    \caption{Distribution of variable importance versus maximum absolute correlation with any truly active predictors in the synthetic data. The $x$-axis is the maximum absolute correlation between each predictor across all five active predictors.}
    \label{fig:vim_corr}
\end{figure}

Figure~\ref{fig:vim_corr} reflects the fact that both UK-PLS and spatial RF allocate the contribution of the true predictors onto others that are highly correlated with them, and conversely, only the predictors that are highly correlated with the true ones were found to have a meaningful contribution in the model. 
This plot also suggests that the greedy tree-based algorithm tends to favor a more parsimonious model when autocorrelation is present among the predictors, {since spatial RF assigns a close-to-zero contribution to most predictors, and only a few are assigned to have high importance. This aligns with our knowledge that if one of the autocorrelated features is selected into a regression tree model, the remaining ones are less likely to further improve model accuracy and thus less likely to enter the model and be identified as important predictors.} The full variable importance plot {leads to the same observations, and} is presented in Appendix~\ref{app:varimp} for completeness.

\subsection{Results for Seattle and National Data}
\label{subsec:data-results}

We examine the proposed variable importance measure at each quartile of each predictor, i.e. $q_1, q_2$ and $q_3$ are $0.25, 0.5$ and $0.75$ respectively. We compare UK-PLS and spatial RF models and look at three contrasts, $\bar\mu_{j,2} - \bar\mu_{j,1}$, $\bar\mu_{j,3} - \bar\mu_{j,2}$ and $\bar\mu_{j,3} - \bar\mu_{j,1}$, to evaluate the contribution of each predictor.

\begin{figure}[!ht]
    \centering
    \includegraphics[width=16.5cm]{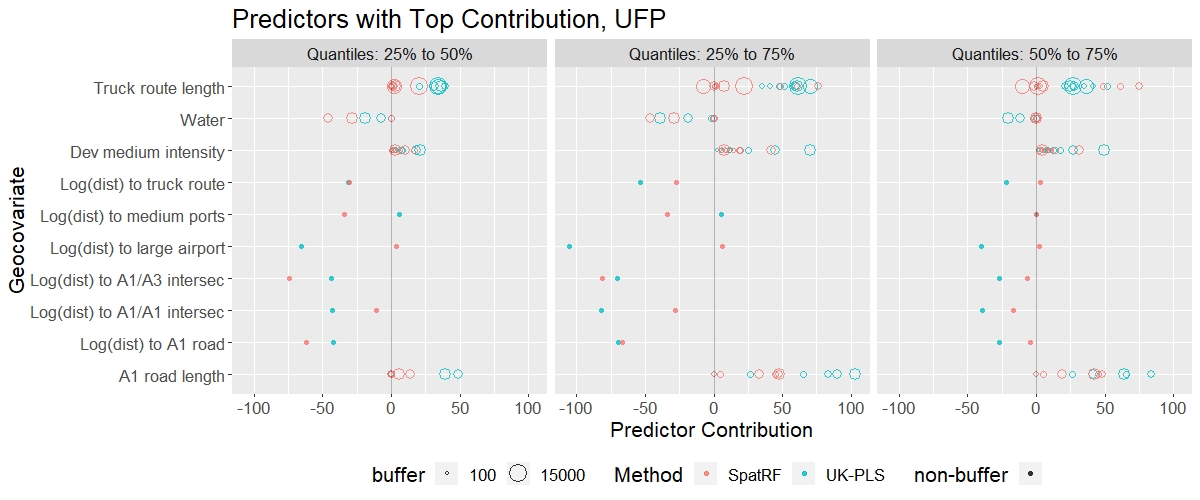}
    \caption{Variable importance plot for the prediction of UFP concentration, showing predictors with top 5 contribution for either method for at least one contrast. All buffer sizes are included if one of them is within the top 5 important predictors.}
    \label{fig:ufp_vip}
\end{figure}

Figure~\ref{fig:ufp_vip} 
visualizes the contribution of predictors having the greatest importance in predicted UFP concentration with the Seattle data. Despite similar predicted maps between UK-PLS and spatial RF, the plots highlight the difference in the mechanisms captured by each model.
In particular, spatial RF identifies the length of truck routes and closeness to major roads as major contributors to predicted UFP concentration in the Seattle TRAP study, while UK-PLS highlights the distance to large airport as a more significant contributor. As known from prior studies, jet engine exhaust is a significant source of UFP \citep[see e.g.][]{hudda2018aviation}, which suggests UK-PLS as a more sensible candidate in terms of scientific interpretation. 
In addition, the UK-PLS model is more consistently influenced by truck traffic and general traffic on large A1 roads than the spatial RF predictions. It is also reasonable that a linear model would perform well in a relatively homogeneous and small area where relationships between sources and pollution levels are consistent across the domain.

{This variable importance measure also reveals the greedy nature of tree building algorithms here.}
{For instance, although both UK-PLS and spatial RF find the length of truck routes within several buffer sizes to be important predictors of UFP concentration, spatial RF highlights only a few of these autocorrelated predictors in contrast to UK-PLS, which highlights all of them, as can be seen in Figure~\ref{fig:ufp_vip}.} Further, the covariate(s) and size of the buffer highlighted varies across quantile contrasts with spatial RF, whereas this is more consistent across quantile contrasts for UK-PLS. This is related to the non-linear property of spatial RF (in contrast to the linear UK-PLS model), namely, the magnitude of effects of the same covariate could differ at different levels of its distribution.

\begin{figure}[!ht]
    \centering
    \includegraphics[width=16.5cm]{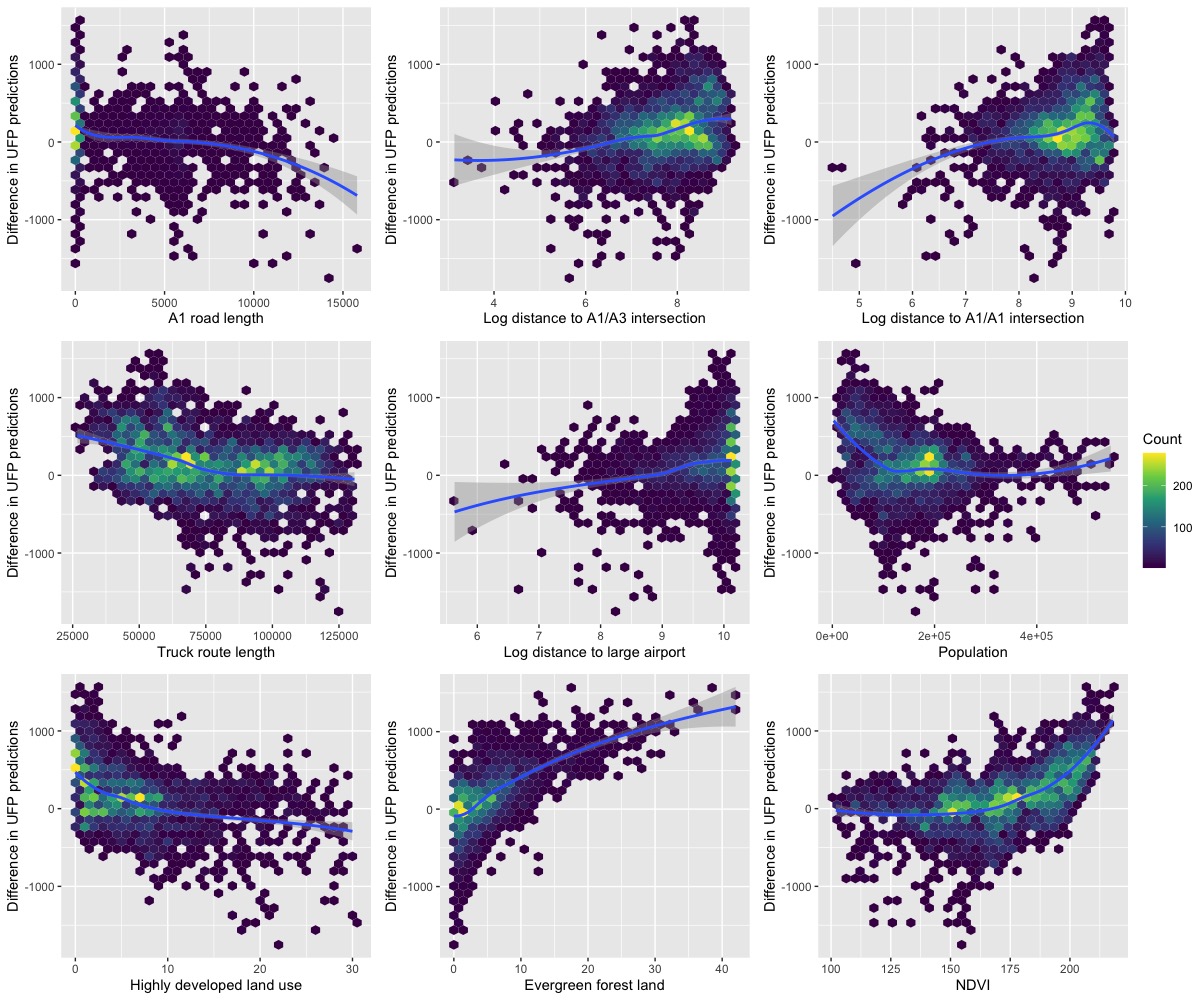}
    \caption{Hexagonal bin plot showing the difference between spatial RF (PL) and UK-PLS (the subtrahend) predictions of UFP concentration on gridded locations, versus the distribution of predictors with the greatest difference in variable importance between models. The color reflects the number of points falling to each small region of the plot. Locally weighted scatterplot smoothing (LOESS) curves are added to show the overall trend.}
    \label{fig:gridufp_hexbin}
\end{figure}

The proposed variable importance measure could also provide insight on how the performance of different fitted models would differ on newly observed data points. 
Recall that for the Seattle TRAP study, we evaluated the trained UK-PLS and Spatial RF (PL) models on an additional set of 2815 gridded locations (with higher resolution within the same study region) that were not used to train the prediction models.
Figure~\ref{fig:gridufp_hexbin} shows how the difference in predicted values between models vary with a set of predictors, for which UK-PLS and spatial RF had the most different variable importance measures as given by the original training data. 
The analysis based on the original training data finds that spatial RF differs from UK-PLS by -4.11 units when looking at the contribution of NDVI (buffer size 5km) changing from its $25\%$ and $50\%$ quantile, and differs by +41.02 units when looking at the change from $50\%$ to $75\%$ quantiles. Therefore, it is expected that at the higher end in the distribution of NDVI we would observe predictions from a spatial RF model would grow faster, or decline slower, compared with UK-PLS, which is indeed reflected in the plot on the bottom right panel. Similar interpretations can be observed in the trend of evergreen forest land (buffer size 3km), highly-developed land use (buffer size 5km) and truck route length (buffer size 10km, albeit less noticeable), among others. There are also signs of greater {variability in the difference between predictions when the distance to A1/A3 intersections, A1/A1 intersections or airports is large, which is consistent with the fact that these predictors were found to play different roles in the two models.}
Figure~\ref{fig:cohortufp_hexbin} in the Appendix presents a similar comparison, visualizing the differences in predictions at the residential locations of an epidemiological cohort throughout the greater Seattle area, as opposed to the gridded locations in Figure~\ref{fig:gridufp_hexbin}. We observe that the difference between UK-PLS and spatial RF is less evident at the cohort locations, which are more spatially aligned with and better represented by the monitoring locations, in contrast to the gridded locations which cover less populated areas as well. This further validates our argument that models with similar behaviors on the training data (e.g., the monitoring sites) could have meaningful differences when extrapolated to new locations (e.g., the gridded locations); and with the aid of our variable importance measure, the latter can be anticipated and captured by a variable importance analysis on just the training data.

\begin{figure}[ht]
    \centering
    \includegraphics[width=16.5cm]{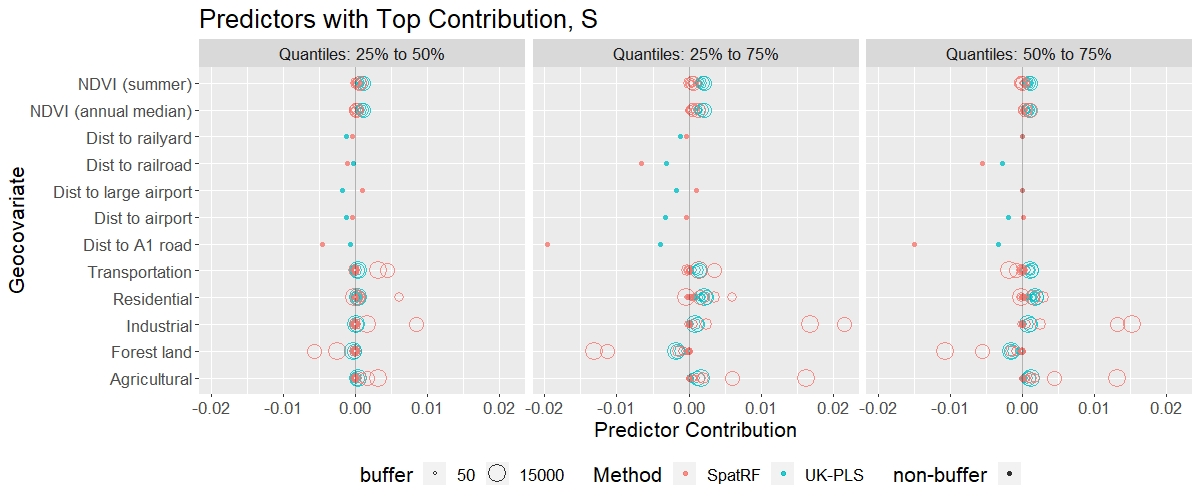}
    \caption{Variable importance plot for the prediction of S concentration, showing predictors with top five contributions for either method for at least one contrast. All buffer sizes are included if one of them is within the top five important predictors.}
    \label{fig:ec_vip}
\end{figure}

In contrast to the relatively homogeneous and small area of Seattle, our analyses on the national data demonstrate the use of variable importance measure when greater spatial heterogeneity in the distribution of pollutant concentrations is present. 
Figure~\ref{fig:ec_vip} reveals that the predictions from spatial RF were greatly driven by proximity to A1 roads and a range of land use features, especially the amount of industrial, agricultural and forest lands within certain buffer sizes. UK-PLS identified a similar set of influential predictors, but all with smaller and more consistent magnitudes across different buffer sizes. While UK-PLS and spatial RF achieved similar prediction accuracy ($R^2$'s of 0.89 and 0.90 respectively), the extreme influence of a few buffer sizes in the spatial RF model raises concerns about generating extreme predicted values (potentially at new, unobserved locations), and also brings challenges to the scientific interpretation.

\section{Discussion}
\label{sec:disc}

{Our investigation starts with the assessment of air pollution exposures in two real-world studies covering a small and large geographic region, respectively.} We apply two machine learning algorithms -- universal kriging with partial least squares and a spatial random forest -- to predict exposure throughout each region.  We compare the results of these two approaches, both in terms of standard prediction model performance summaries and a new variable importance metric proposed in this paper.  
We found in the small geographic region setting, that although the two models had comparable prediction performances, the variable importance metric indicated that some geographic features, specifically distance to large airport, distance to main roads, and length of truck routes, were differently important contributors to the predictions of UFP concentration from UK-PLS and spatial RF. 
When a larger geographic region is of interest, we observed UK-PLS and spatial RF both identifying distances to main roads and the amount of different land use as contributors to the prediction of Sulfur concentration, while a few buffer sizes for land use features were assigned extreme influence by spatial RF. 
Our analysis of the synthetic data showed that the importance assigned by the proposed measure on each predictor was positively associated with its correlation with the truly active predictors, where spatial RF favored a more parsimonious model with larger magnitude of contributions from each predictor comparing to UK-PLS. 
Given that epidemiologic cohort studies rely on predicted air pollution exposures for making inference about health effects, use of this variable importance metric has the potential to allow new insights into how seemingly similar exposure metrics may lead to different inferences.    

A primary motivation for our work is to improve exposure assessment for air pollution {cohort} studies where the pollution exposure surface is predicted from a spatial model trained on monitoring data. Recent developments in sensor technology, monitoring study design, and statistical modeling methods have made it possible to construct accurate exposure models for a variety of pollutants at both local and national scales, as illustrated by the data we analyzed in this paper. This state of affairs introduces a new set of challenges since many choices need to made in designing a particular exposure model, and in some cases, there are already multiple published models to choose from with overlapping spatial and temporal domains. For example, at least three research groups have developed models for PM\textsubscript{2.5} that cover large portions of the United States \citep{yanosky2009predicting, di2016assessing, kirwa2021fine}.

A typical strategy is to select the model with the smallest out-of-sample prediction error (or highest $R^2$) as a way of minimizing exposure measurement error. This is generally a sound strategy, although it is now known that the model with the highest $R^2$ does not always lead to the best health effect inference \citep{szpiro2011does}, in part owing to the complexity of balancing different types of measurement error and interactions between measurement error and covariates in the health model \citep{szpiro2013measurement, cefalu2014does, bergen2016multipollutant}. Given this context, it makes sense to utilize variable importance as we have developed it here as an additional tool to decide between models, giving primacy to those models that are more interpretable in terms of what is known about sources and dispersion of the air pollutant being modeled. An open question that we will consider in future research is how to balance prediction accuracy and variable importance in selecting an exposure model, e.g., when would there be a large enough difference in model performance across modeling approaches that would lead us to focus almost exclusively or entirely on model performance and not put any weight on variable importance? 

In some air pollution epidemiology studies, the specific pollutant used as the exposure is regarded as a marker for source-specific pollution. For example, many studies have utilized elemental carbon (EC), black carbon (BC), oxides of nitrogen (NO, NO\textsubscript{2}), and fine particulate matter (PM\textsubscript{2.5}) as markers of traffic-related pollution (TRAP). The strength with which findings about these pollutants implicate traffic as a pollutant source depends on how much of a role traffic played in the exposure model. In a recent overview of health effects of TRAP on a wide variety of health outcomes, systematic but ad-hoc methods were used to determine which exposure models could be regarded as sufficiently traffic-specific \citep{boogaard2022long}, and the selection process would have benefited from availability of a variable importance metric like ours that quantifies the contribution of traffic-related covariates to the predicted concentrations.


The variable importance measure we present is flexible, intuitive, and generally applicable to machine learning models that account for spatial correlation. This leave-one-out approach can be applied to additive models with separable mean and correlation components, including non-linear, ensemble and/or doubly stochastic spatial models. It provides a unifying notion of variable importance which would otherwise be less comparable between different modeling approaches, and we have demonstrated that meaningful differences in the model structure could be found even for models producing similar predictions. 
An informative variable importance measure as ours also facilities deeper understanding of complex prediction models in the methodological aspect: our Seattle and national data examples illustrate the greedy nature of tree building algorithms which is already well-known; but such information would not be straightforward to obtain otherwise for more complex black-box models.

{Our approach is an example of extrinsic variable importance measure, which is intimately tied to the specific prediction model, see e.g. \citet{breiman2001random, strobl2007bias}; on the contrary, intrinsic variable importance is model-agnostic and corresponds to the best possible model (which is often unknown) within a certain class \citep{van2006statistical, lei2018distribution, williamson2021nonparametric}. Both types of variable importance measures are meaningful depending on the practical use cases, and extrinsic metrics are useful especially for the interpretation and selection of models.}



One interesting extension of our approach would be to estimate the uncertainty of variable importance measures. As a simple and naive solution, sample splitting such as cross-validation or bootstrap can both provide an uncertainty estimate, though more careful treatment is needed with the existence of spatial correlation. Also, as is the case for many variable importance measures, the proposed approach reflects the association between predictors and the outcome captured by a given model, rather than causal effects. And consequently, when autocorrelation between predictors is present and only a few of them are truly contributing to the outcome, it could be challenging to disentangle them. {However, our analysis of synthetic data indicates that only the predictors that are highly correlated with the truly active ones are likely to be identified as important by the proposed measure.}

\bibliographystyle{rss}
\bibliography{ref}

\newpage
\clearpage
\appendix
\pagenumbering{arabic}

\begin{center}
{\large\bf APPENDIX}
\end{center}

\section{Annual Average Pollutant Concentration and Prediction Results}
\label{app:annavg}
\subsection{Data Description}
\begin{figure}[!ht]
    \centering
    \includegraphics[width=16.5cm]{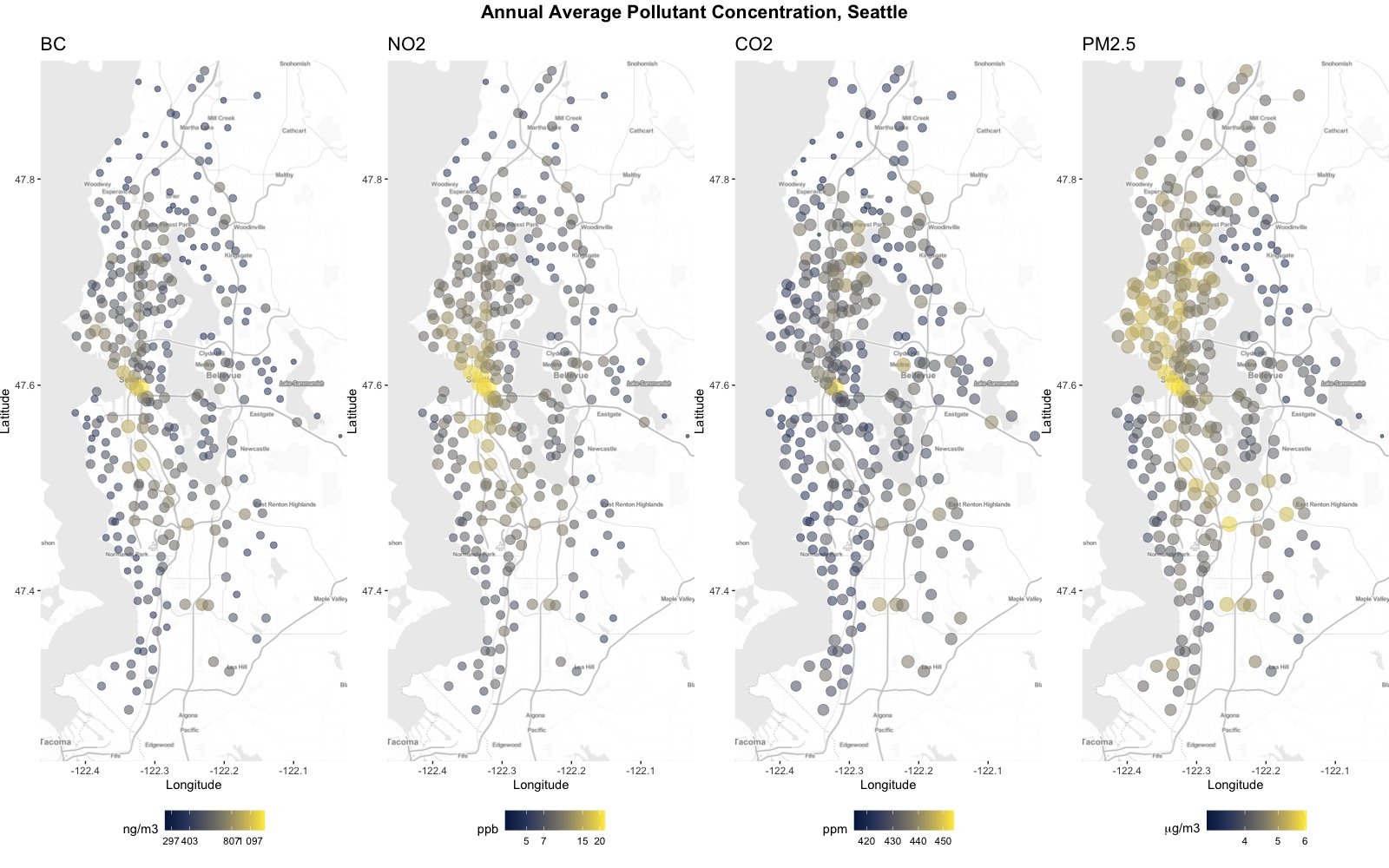}
    \caption{Annual average concentration of BC, NO\textsubscript{2}, CO\textsubscript{2} and PM\textsubscript{2.5} at mobile monitoring locations in the Seattle dataset}
    \label{fig:app-annavg-sea}
\end{figure}

\begin{figure}[!ht]
    \centering
    \includegraphics[width=16.5cm]{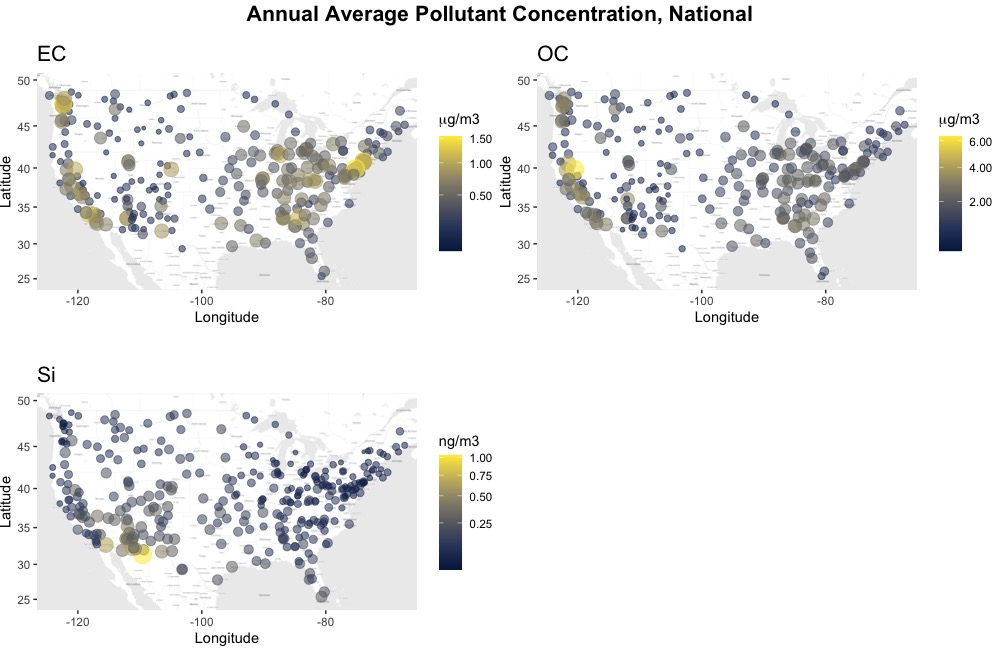}
    \caption{Annual average concentration of EC, OC and Si at monitoring locations in the national dataset}
    \label{fig:app-annavg-national}
\end{figure}

This section presents the distributions of pollutants that are not discussed in detail in the main text.
Figure~\ref{fig:app-annavg-sea} visualizes the estimated annual average concentration of BC, NO\textsubscript{2}, CO\textsubscript{2} and PM\textsubscript{2.5} in the Seattle TRAP study, and Figure~\ref{fig:app-annavg-national} presents the annual average concentration of OC, S and Si in the national data.

\subsection{Synthetic Data Description}
\begin{figure}
    \centering
    \includegraphics[width=17cm]{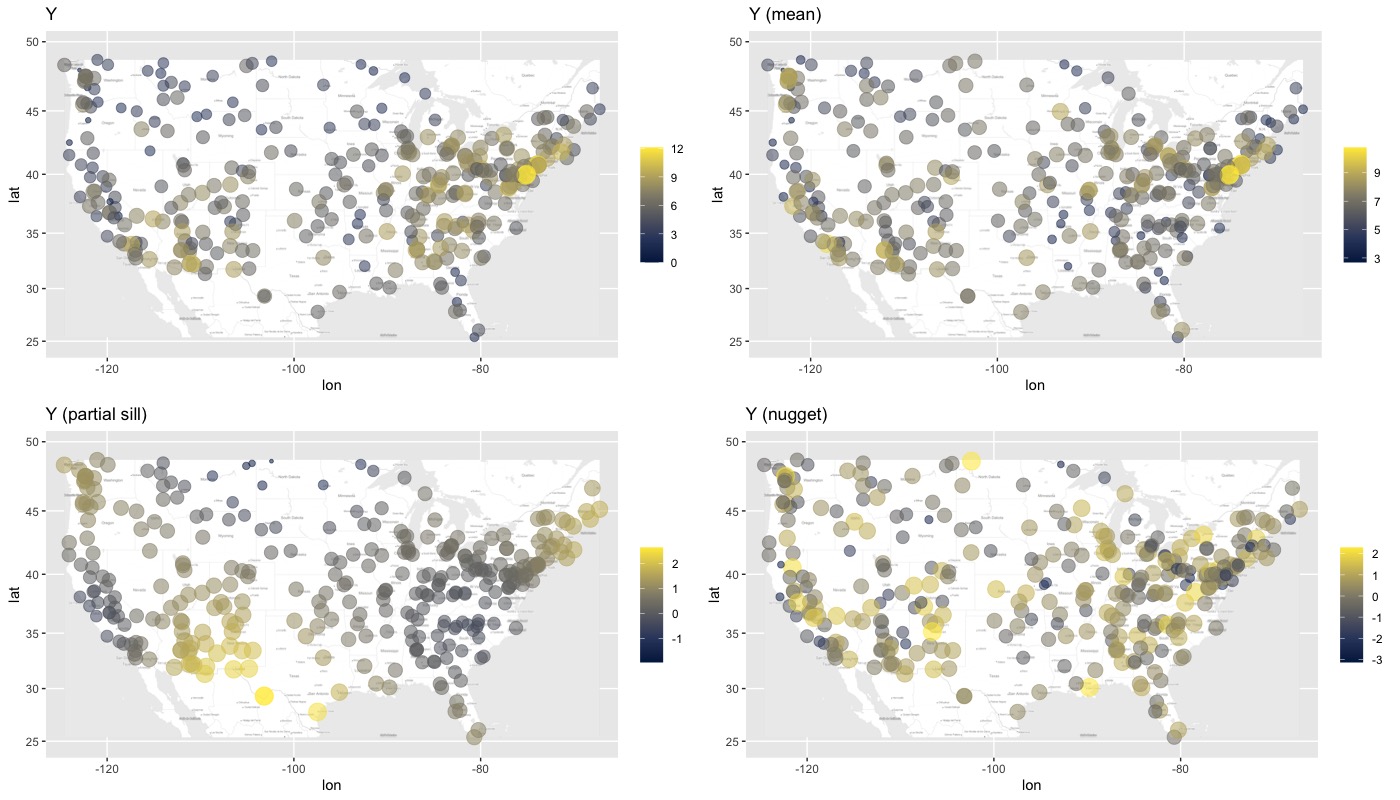}
    \caption{Decomposition of the synthetic outcome}
    \label{fig:y_decomp}
\end{figure}

{Figure~\ref{fig:y_decomp} visualizes the overall distribution and decomposition of the synthetic data, i.e. variability coming from the mean, partial sill and nugget, respectively.}

\subsection{Prediction Models}
In addition to our primary models, UK-PLS and spatial RF-PL, we also investigated the performance of {spatial RF with nonparametric optimization approach \citep[SpatRF-NP, see][]{wai2020random} along with} four benchmark models as a comparison:

\begin{itemize}
    \item RF: random forest implemented by the \texttt{randomForest} R package ignoring spatial correlation;
    \item TPRS: spatial smoothing via thin plate regression splines implemented by the \texttt{mgcv} R package;
    \item RF-TPRS: a two-step procedure that first runs RF, and then conducts TPRS spatial smoothing on the residuals from RF;
    \item TPRS-RF: a two-step procedure that first runs TPRS, and then applies RF on the residuals from TPRS.
\end{itemize}

\begin{table}[ht]
\centering
\begin{tabular}{@{}llllllll@{}}
\toprule
      & UK-PLS & RF   & TPRS & RF-TPRS & TPRS-RF & SpatRF (PL) & SpatRF (NP) \\ \midrule
UFP   & 0.81   & 0.75 & 0.76 & 0.79    & 0.80    & 0.78        & 0.78        \\
BC    & 0.65   & 0.60 & 0.57 & 0.67    & 0.64    & 0.67        & 0.67        \\
NO\textsubscript{2}   & 0.77   & 0.70 & 0.70 & 0.76    & 0.74    & 0.75        & 0.74        \\
CO\textsubscript{2}   & 0.56   & 0.47 & 0.44 & 0.57    & 0.55    & 0.56        & 0.54        \\
PM\textsubscript{2.5} & 0.76   & 0.66 & 0.73 & 0.72    & 0.73    & 0.74        & 0.71        \\ \bottomrule
\end{tabular}
\caption{Cross-validated $R^2$ for each method on the Seattle TRAP data}
\label{tab:sea-r2}
\end{table}

\begin{figure}
    \centering
    \includegraphics[width=16.5cm]{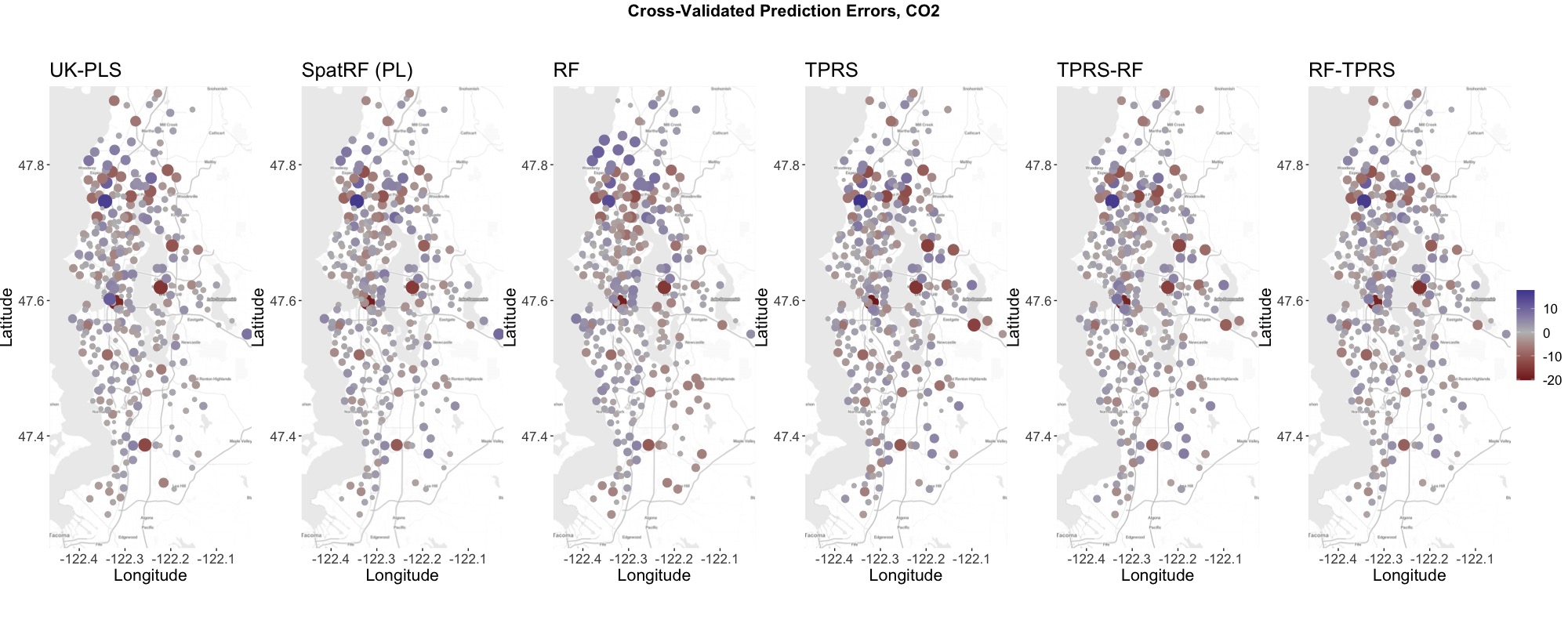}
    \includegraphics[width=16.5cm]{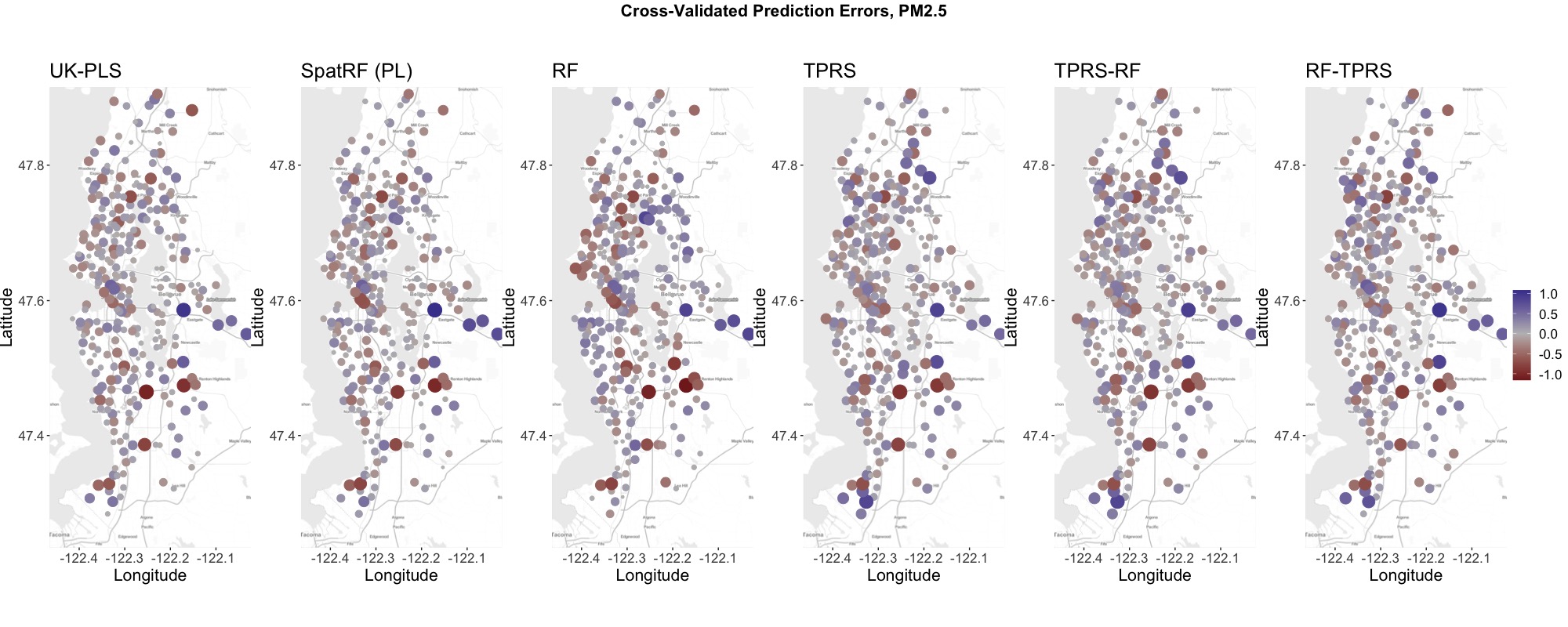}
    \caption{Prediction errors for all pollutants with all models for the Seattle dataset}
    \label{fig:app-sea-cverr}
\end{figure}

Table~\ref{tab:sea-r2} summarizes the cross-validated $R^2$ for all models and all pollutants on the Seattle data. The performance of different models relative to each other reveals different sources of heterogeneity in pollutant concentration: for UFP and NO\textsubscript{2}, purely covariate and spatial effects both account for part of the spatial heterogeneity, reflected by reasonable performance of RF or TPRS alone; accounting for both of them together either in a joint or two-step manner further leads to increased accuracy. The case is similar for BC and CO\textsubscript{2}, where covariate effects appear to be more discernible than spatial effects. PM\textsubscript{2.5}, on the other hand, illustrates a scenario where spatial smoothing alone captures the major source of heterogeneity. UK-PLS and spatial RF have the best overall performance for all pollutants, while neither shows clearly better or worse accuracy than the other. Figure~\ref{fig:app-sea-cverr} shows the cross-validated prediction errors for all models and all pollutants in the Seattle dataset.

\begin{figure}
    \centering
    \includegraphics[width=16.5cm]{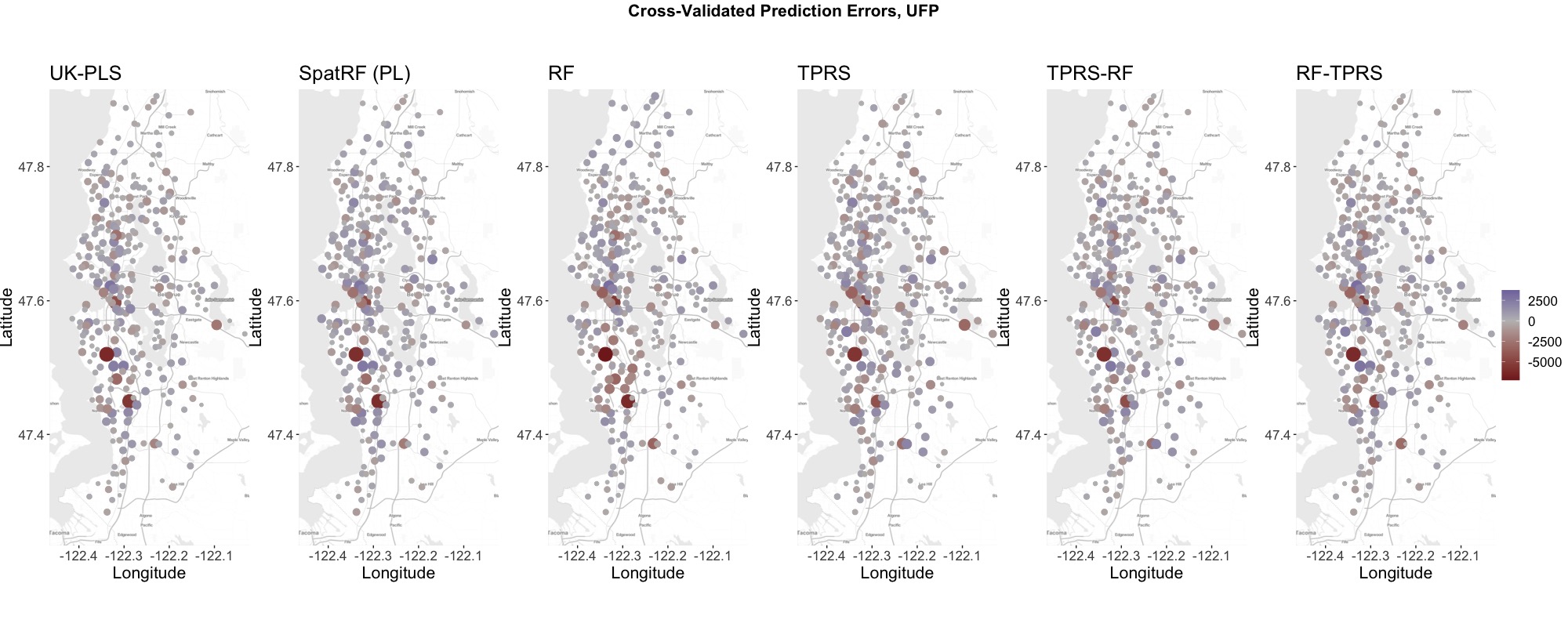}
    \includegraphics[width=16.5cm]{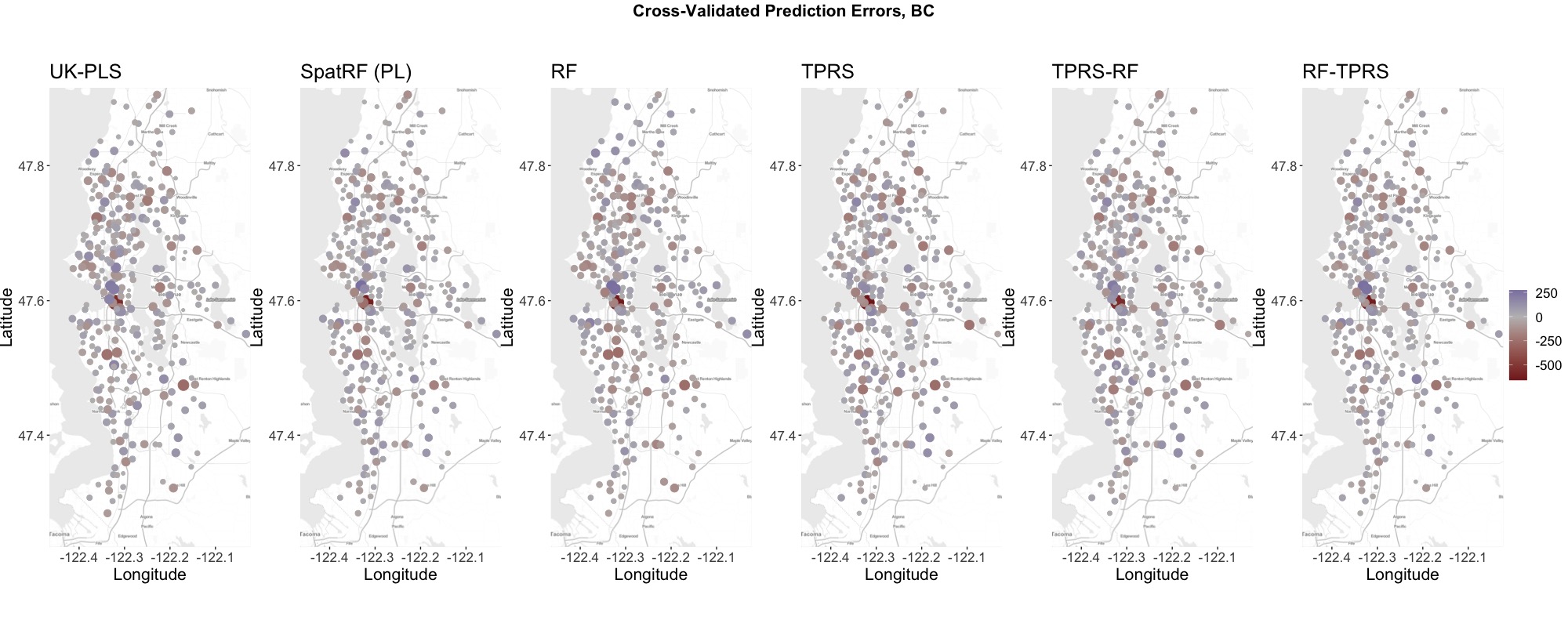}
    \includegraphics[width=16.5cm]{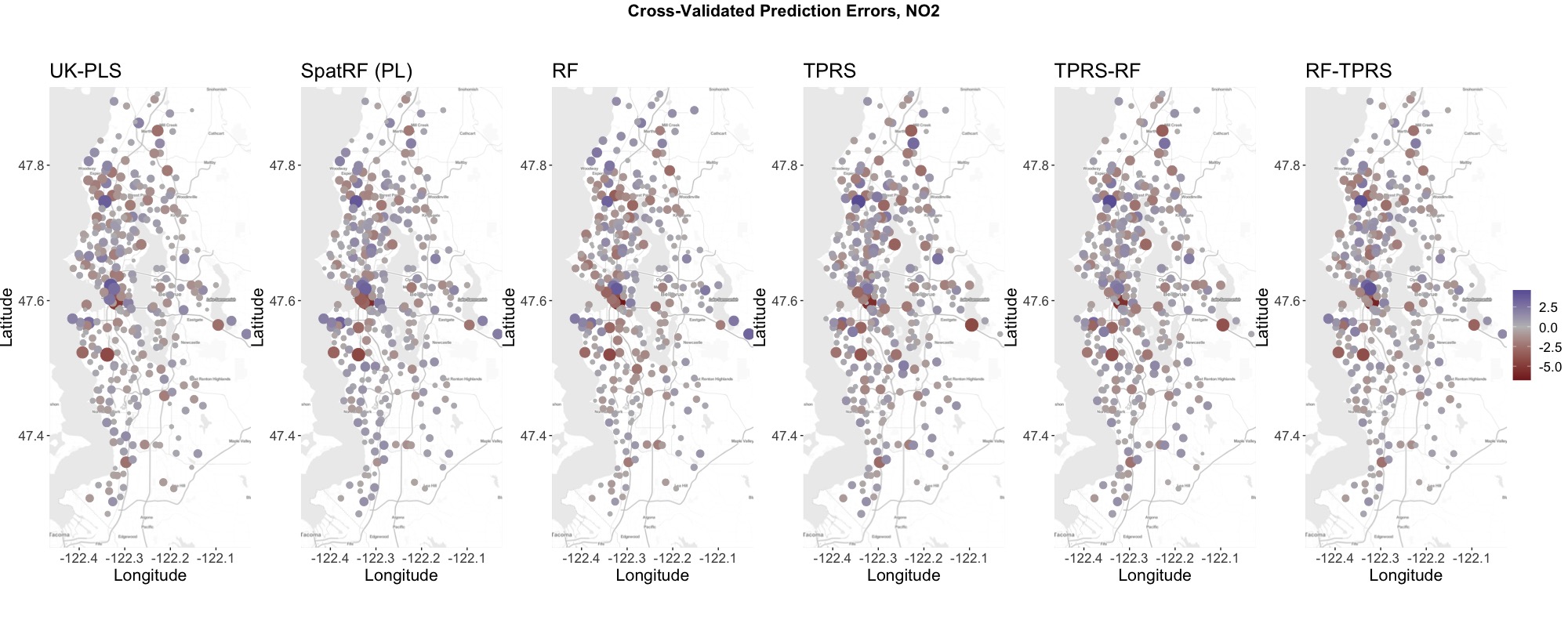}
\end{figure}

Table~\ref{tab:national-r2} compares the predictive performance of each model on the national PM\textsubscript{2.5} sub-species data. Such comparison reflects various scenarios under which different sources of heterogeneity best explain the distribution of outcomes. For EC and OC, the majority of variability comes from covariate effects, as indicated by the poor performance of spatial smoothing (TPRS) alone, while non-linear effects (as captured by RF and spatial RF) are more evident for EC. On the contrary, the spatial component captures a considerable amount of variability for Si and S, and RF has the worst performance on them. This agrees with the findings in \citet{bergen2013national} where adding universal kriging on top of PLS leads to clearly improved accuracy. Figure~\ref{fig:national_cverr} shows the cross-validated prediction errors at all study sites for each method.

\begin{table}[ht]
\begin{tabular}{llllllll}
\hline
   & UK-PLS & RF   & TPRS & RF-TPRS & TPRS-RF & SpatRF (PL) & SpatRF (NP) \\ \hline
EC & 0.72   & 0.79 & 0.23 & 0.81    & 0.73    & 0.82        & 0.81        \\
OC & 0.59   & 0.54 & 0.23 & 0.64    & 0.57    & 0.62        & 0.59        \\
Si & 0.53   & 0.41 & 0.56 & 0.52    & 0.57    & 0.55        & 0.55        \\
S  & 0.89   & 0.76 & 0.93 & 0.89    & 0.94    & 0.90        & 0.87        \\ \hline
\end{tabular}
\caption{Cross-validated $R^2$ for each method on the national PM\textsubscript{2.5} sub-species data}
\label{tab:national-r2}
\end{table}

\begin{figure}
    \centering
    \includegraphics[width=16.5cm]{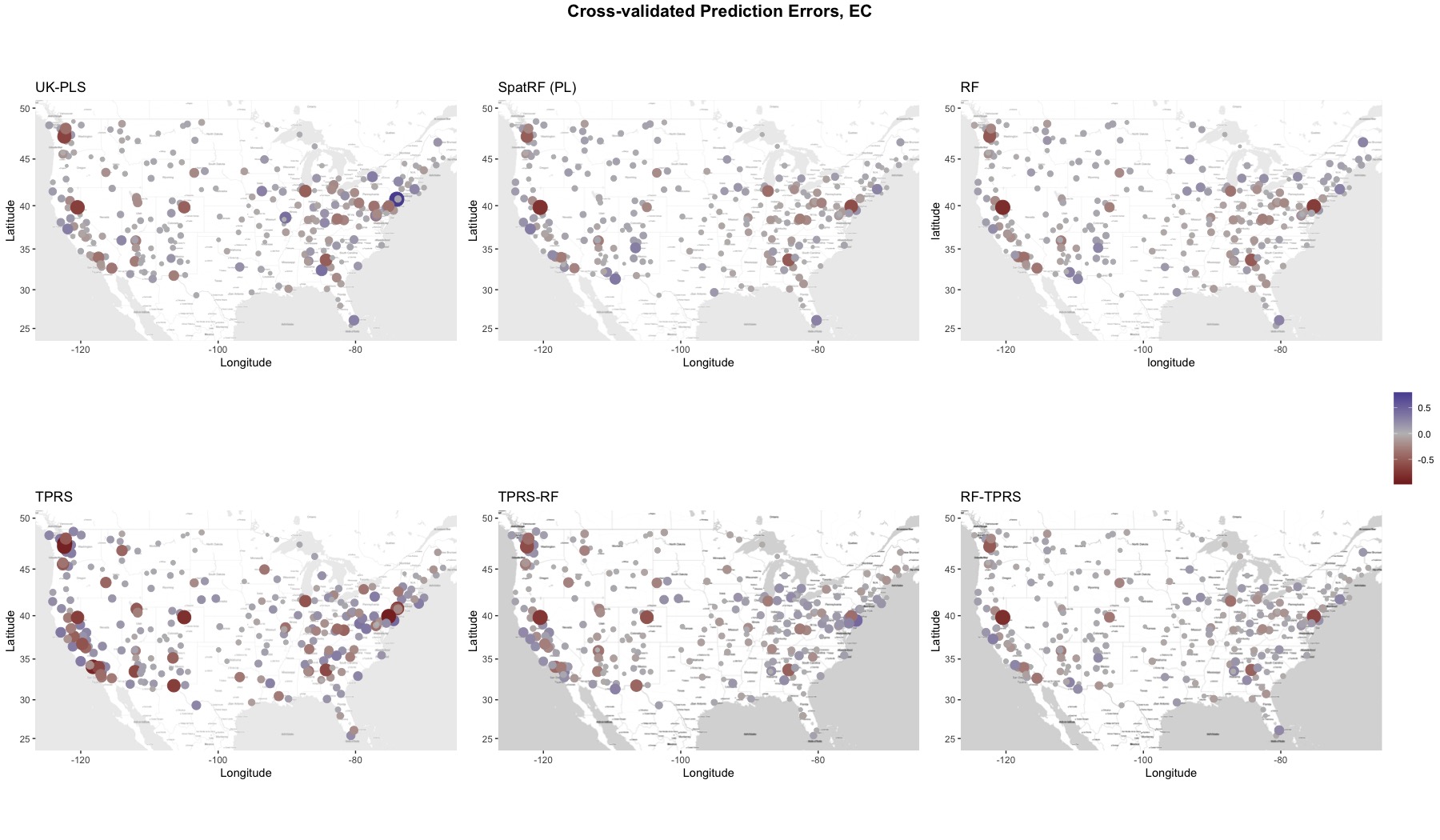}
    \includegraphics[width=16.5cm]{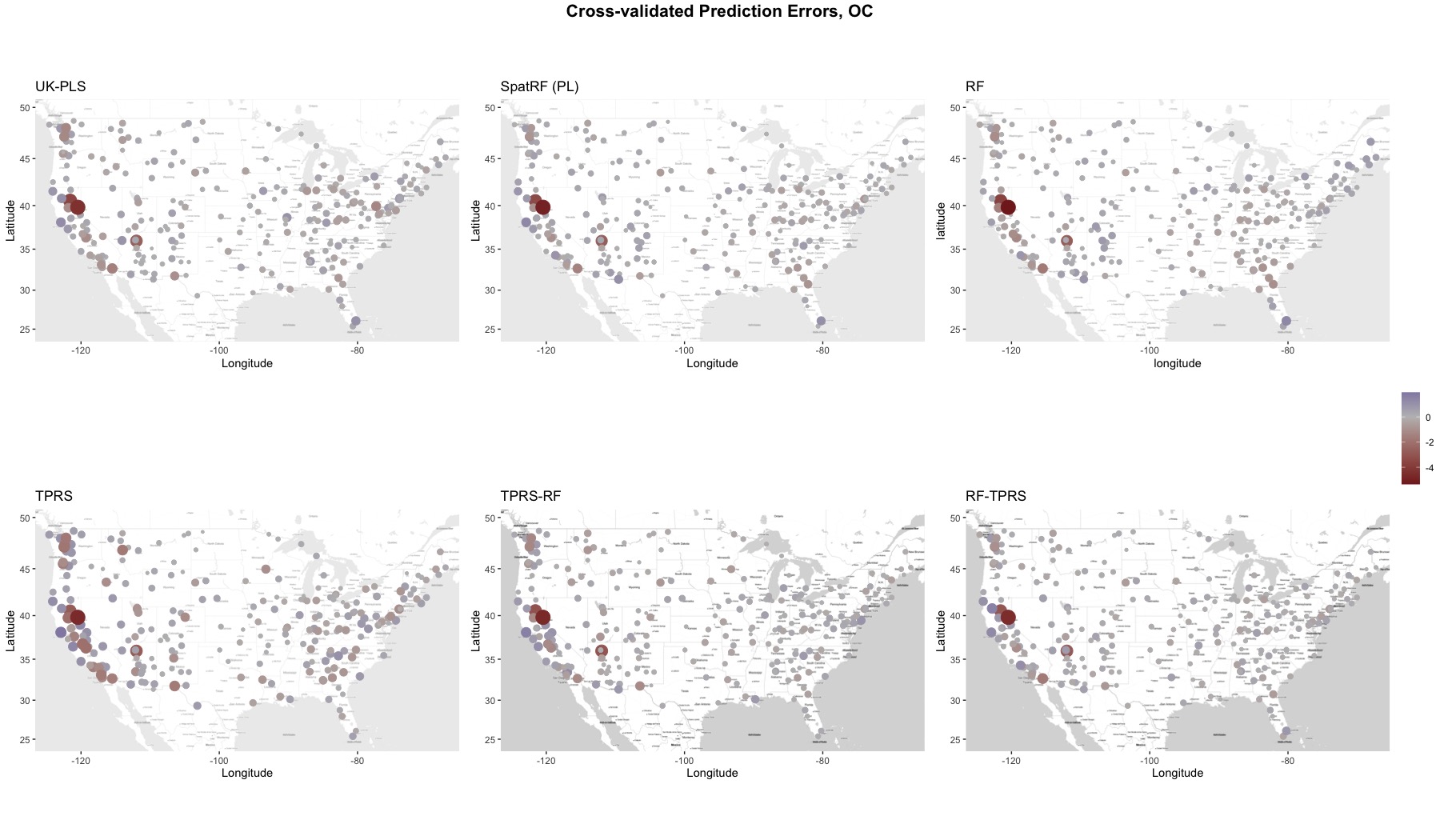}
\end{figure}

\begin{figure}
    \centering
    \includegraphics[width=16.5cm]{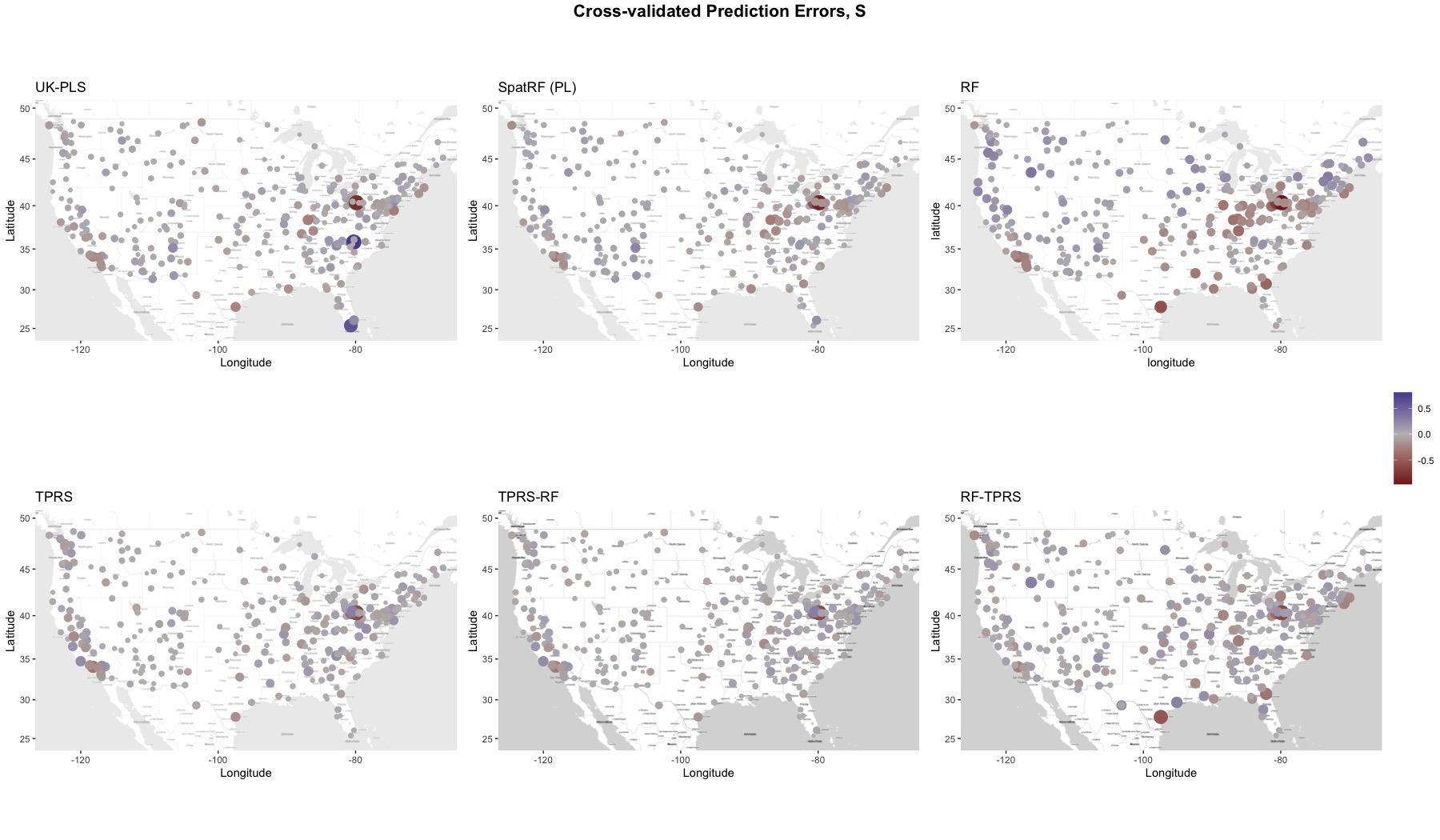}
    \includegraphics[width=16.5cm]{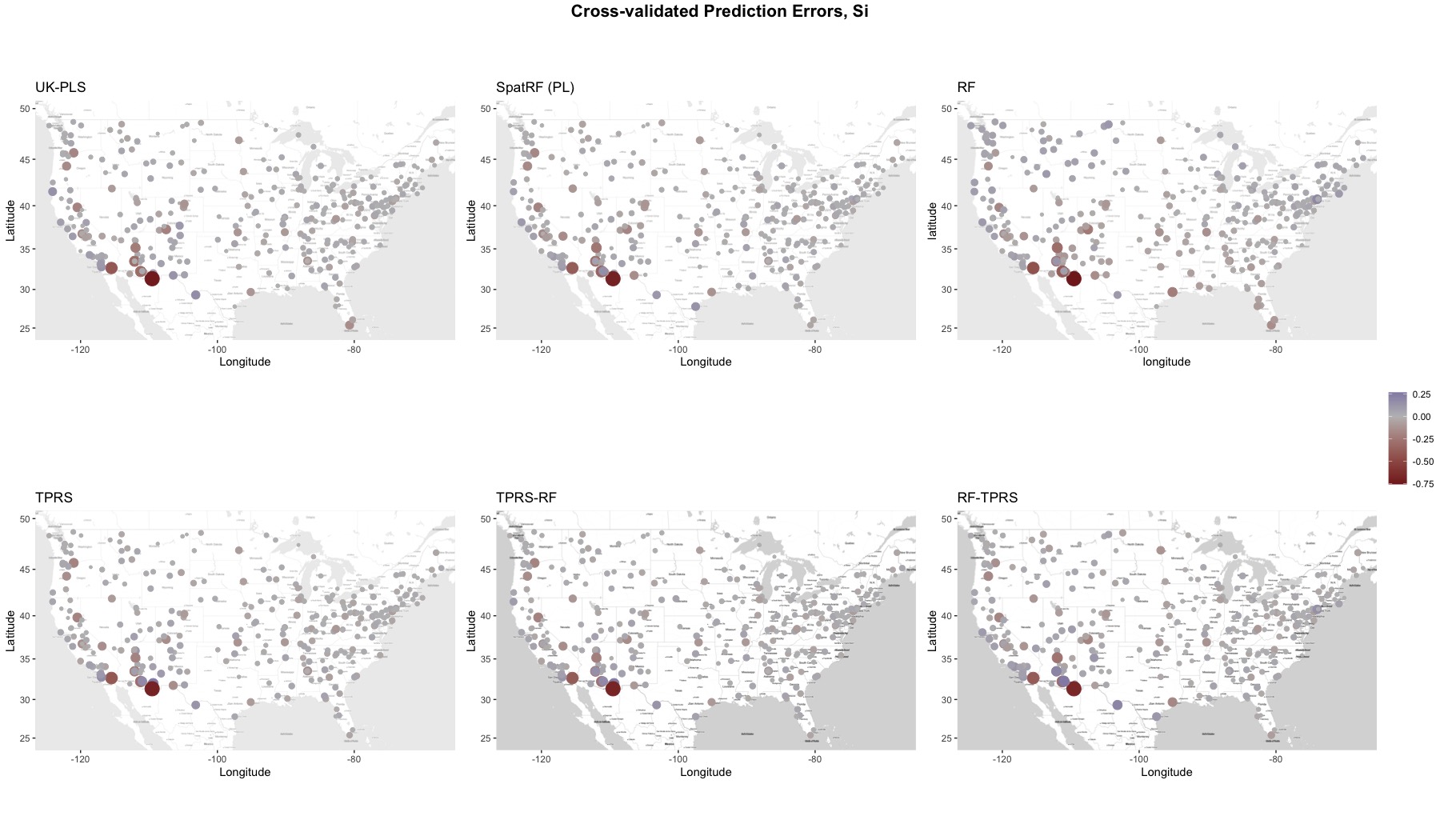}
    \caption{Prediction errors for all pollutents with all models for the national dataset}
    \label{fig:app-national-cverr}
\end{figure}

Table~\ref{tab:r2_synthetic} presents the prediction $R^2$ of different models on the synthetic data, where we observe that models capturing the mean (RF) or correlation (TPRS) only have the lowest accuracy, and UK-PLS which only captures linear relationship has worse performance comparing to more flexible models (two-step models or Spatial RF).

\begin{table}[ht]
\centering
\begin{tabular}{@{}lllllll@{}}
\toprule
UK-PLS & RF   & TPRS & RF-TPRS & TPRS-RF & SpatRF (PL) & SpatRF (NP) \\ \midrule
0.62   & 0.61 & 0.32 & 0.69    & 0.67    & 0.72        & 0.72        \\ \bottomrule
\end{tabular}
\caption{Cross-validated $R^2$ for each method on synthetic data}
\label{tab:r2_synthetic}
\end{table}

\section{Variable Importance Analyses}
\label{app:varimp}

\begin{figure}[!h]
    \centering
    \includegraphics[width=16.5cm]{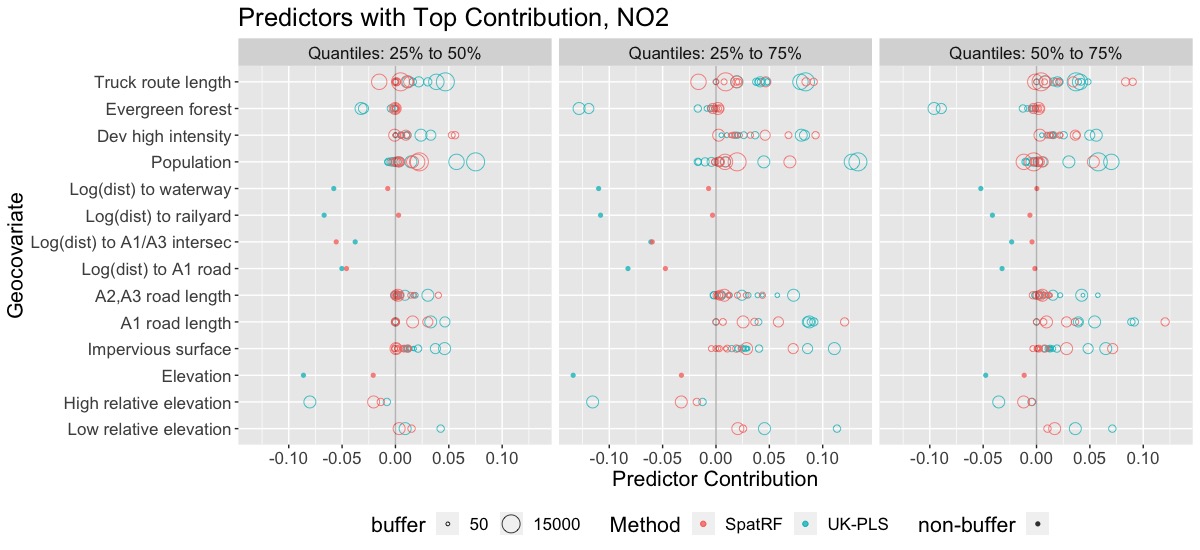}
    \includegraphics[width=16.5cm]{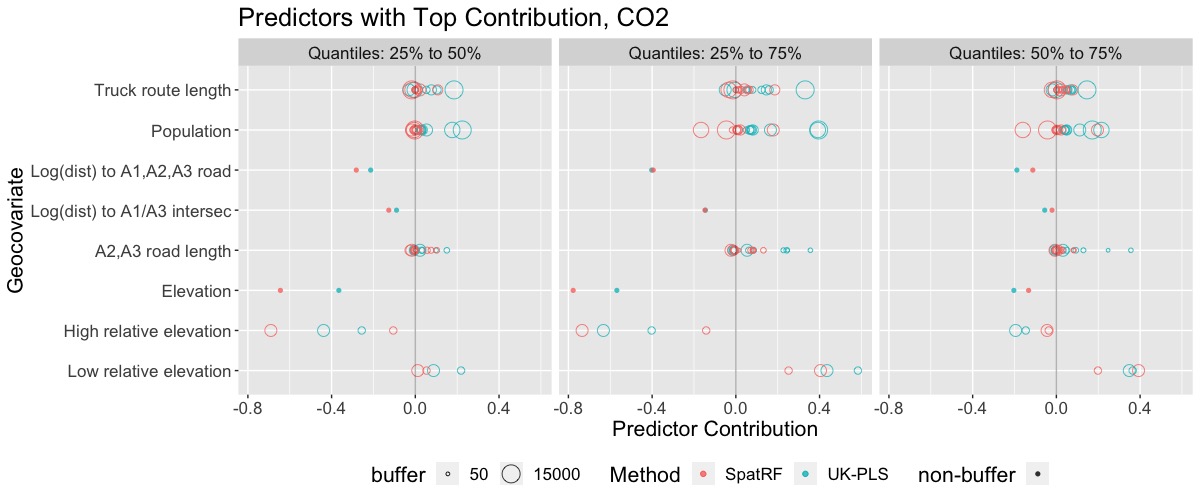}
    \includegraphics[width=16.5cm]{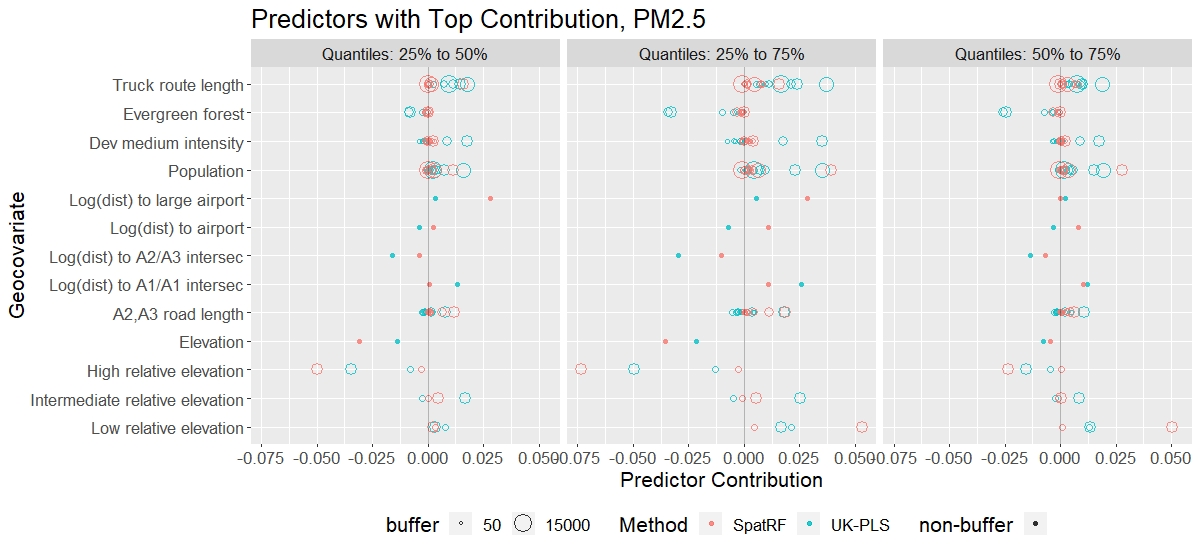}
    \caption{Variable importance plot for the prediction of BC, NO\textsubscript{2},  CO\textsubscript{2} and PM\textsubscript{2.5} concentrations in the Seattle data, showing predictors with top 5 contribution for either method for at least one contrast.}
    \label{fig:app-sea-vip}
\end{figure}

\begin{figure}[ht]
    \centering
    \includegraphics[width=16.5cm]{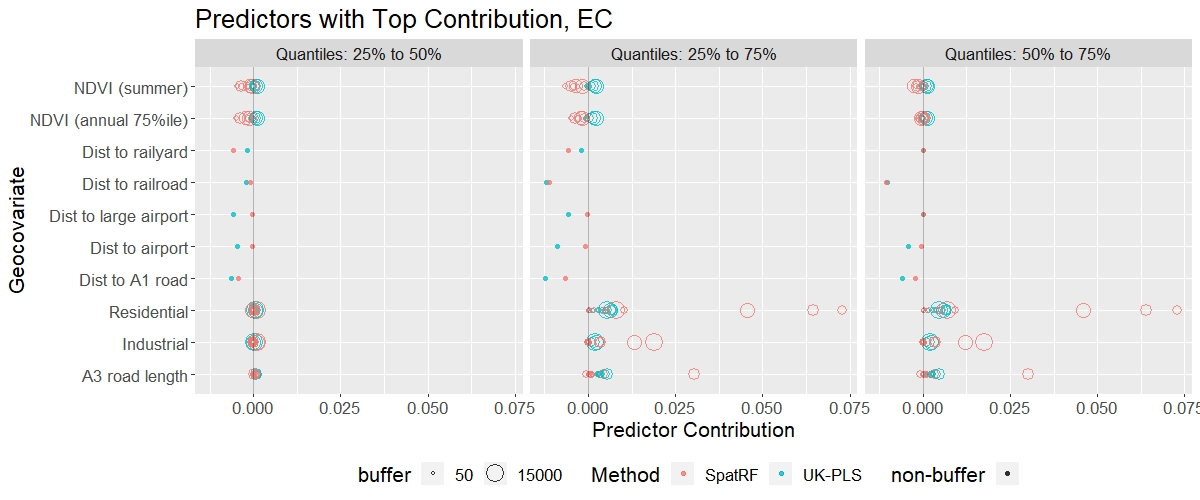}
    \includegraphics[width=16.5cm]{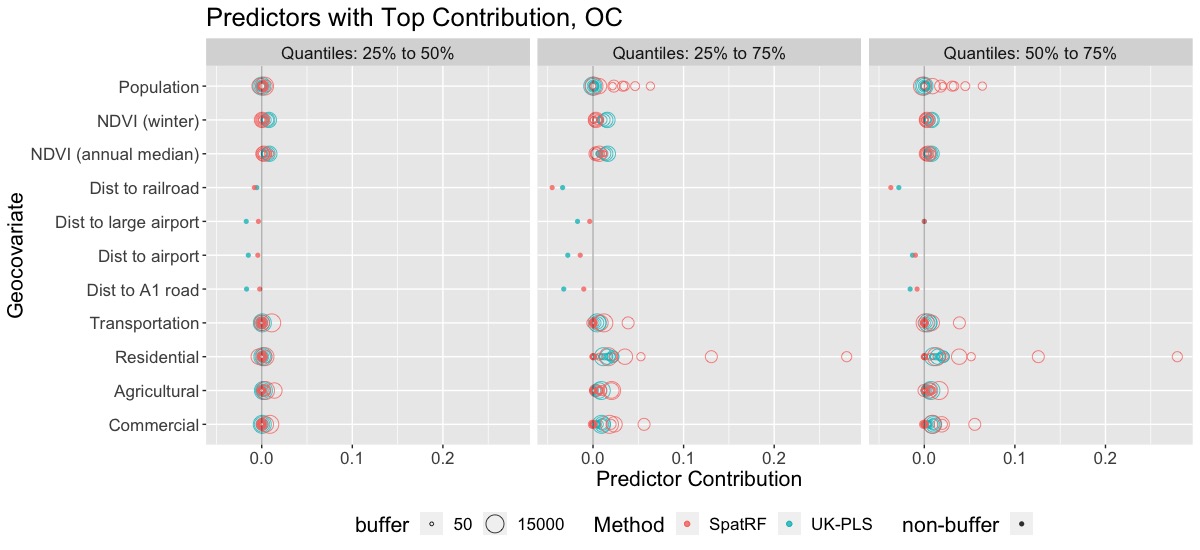}
    \includegraphics[width=16.5cm]{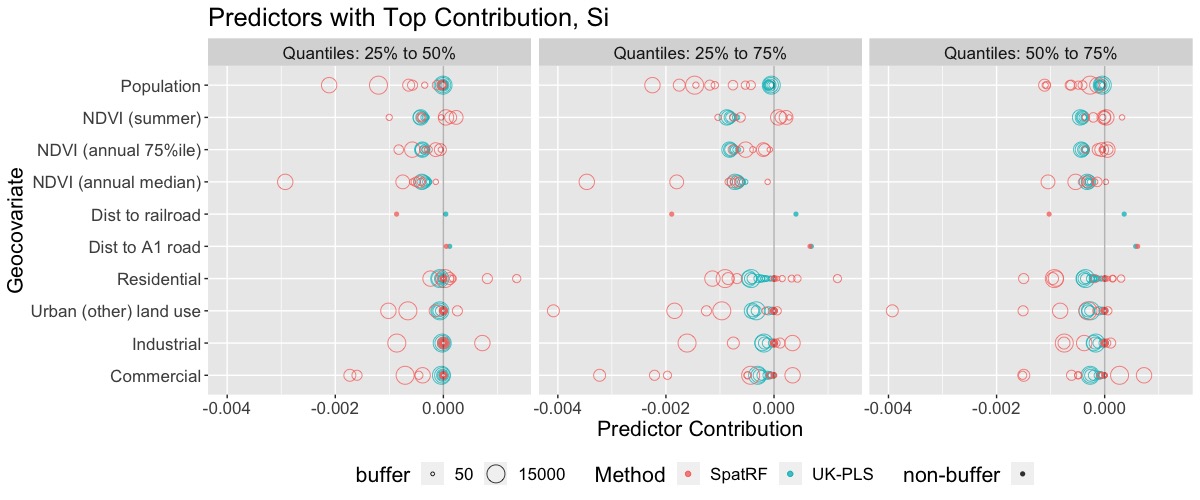}
    \caption{Variable importance plot for the prediction of EC, OC and Si concentration in the national data, showing predictors with top 5 contribution for either method for at least one contrast. }
    \label{fig:app-national-vip}
\end{figure}

Figures~\ref{fig:app-sea-vip} and~\ref{fig:app-national-vip} present our full variable importance results, for all pollutants and all models in the Seattle and national studies, respectively.

\begin{figure}
\centering
    \includegraphics[width=17.5cm]{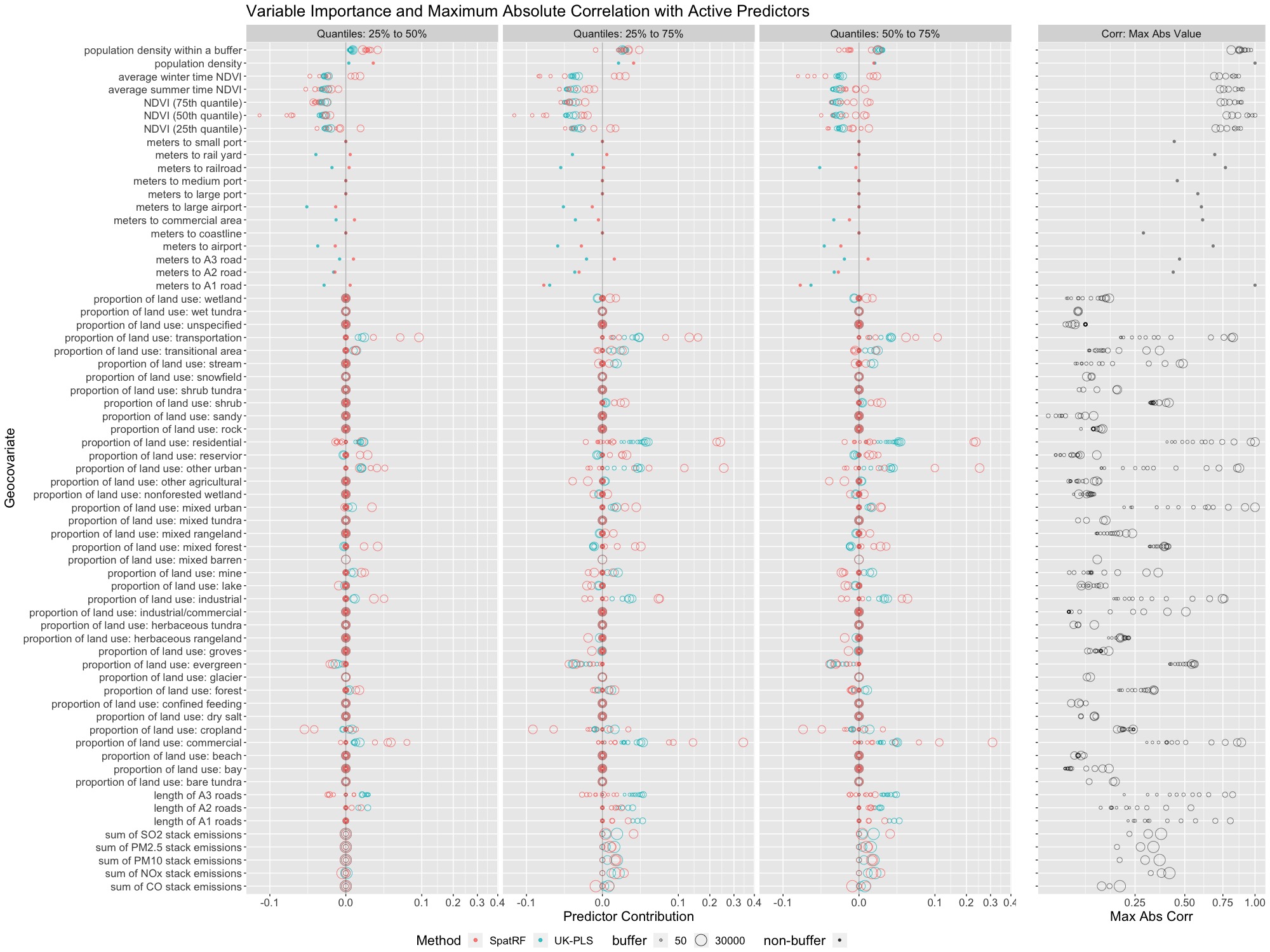}
    \caption{The full variable importance plot for the synthetic data, along with correlation between each predictor and the active predictors. First three columns: variable importance of spatial RF and UK-PLS; last column: maximum absolute correlation between each predictor and the 5 truly active predictors}
\label{fig:varimp_syn}
\end{figure}

Figure~\ref{fig:varimp_syn} visualizes the proposed variable importance measure together with the maximum absolute correlation between each predictor and each truly active predictor, for the synthetic data.

Due to the autocorrelation between predictors, it is unlikely that any variable importance measure would exactly recover the true predictor contributions. 
Instead, a reasonably good measure would highlight predictors that are highly correlated the truly active ones, which would explain the mechanism well enough for practical purposes such as prediction. Our method achieves this, as reflected by the observation that predictors found to have high contribution in the first three panels are either the true ones (e.g. annual median NDVI) or highly correlated with at least one truly active predictor (e.g. transportation land use, which has a maximum absolute correlation above 0.75 with the true predictors as seen from the last panel).
Through this variable importance measure, both UK-PLS and Spatial RF correctly identified the truly active predictors, despite lower magnitude due to autocorrelation between predictors. 

\begin{figure}[!h]
    \centering
    \includegraphics[width=16.5cm]{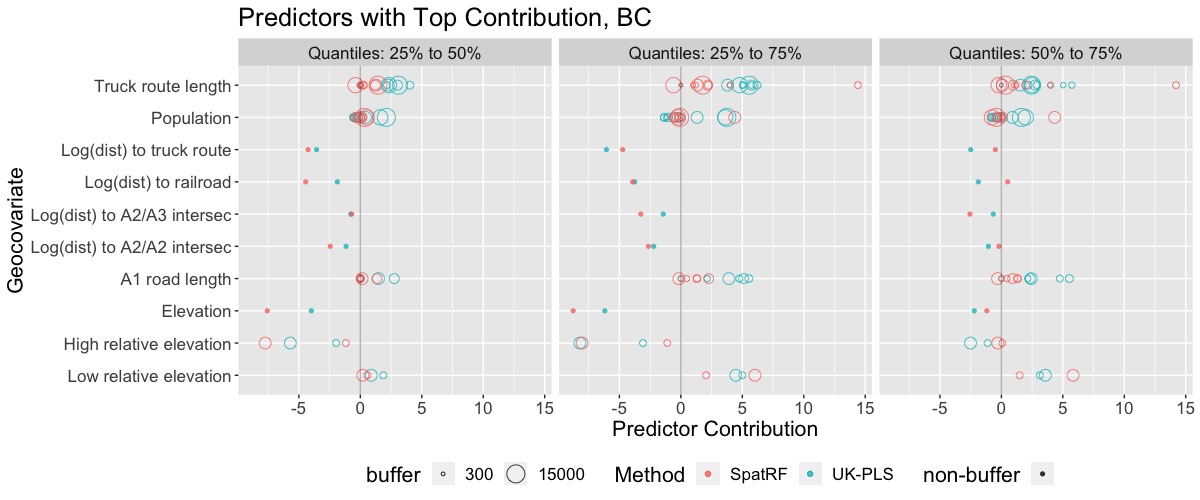}
\end{figure}

\begin{figure}[!ht]
    \centering
    \includegraphics[width=16.5cm]{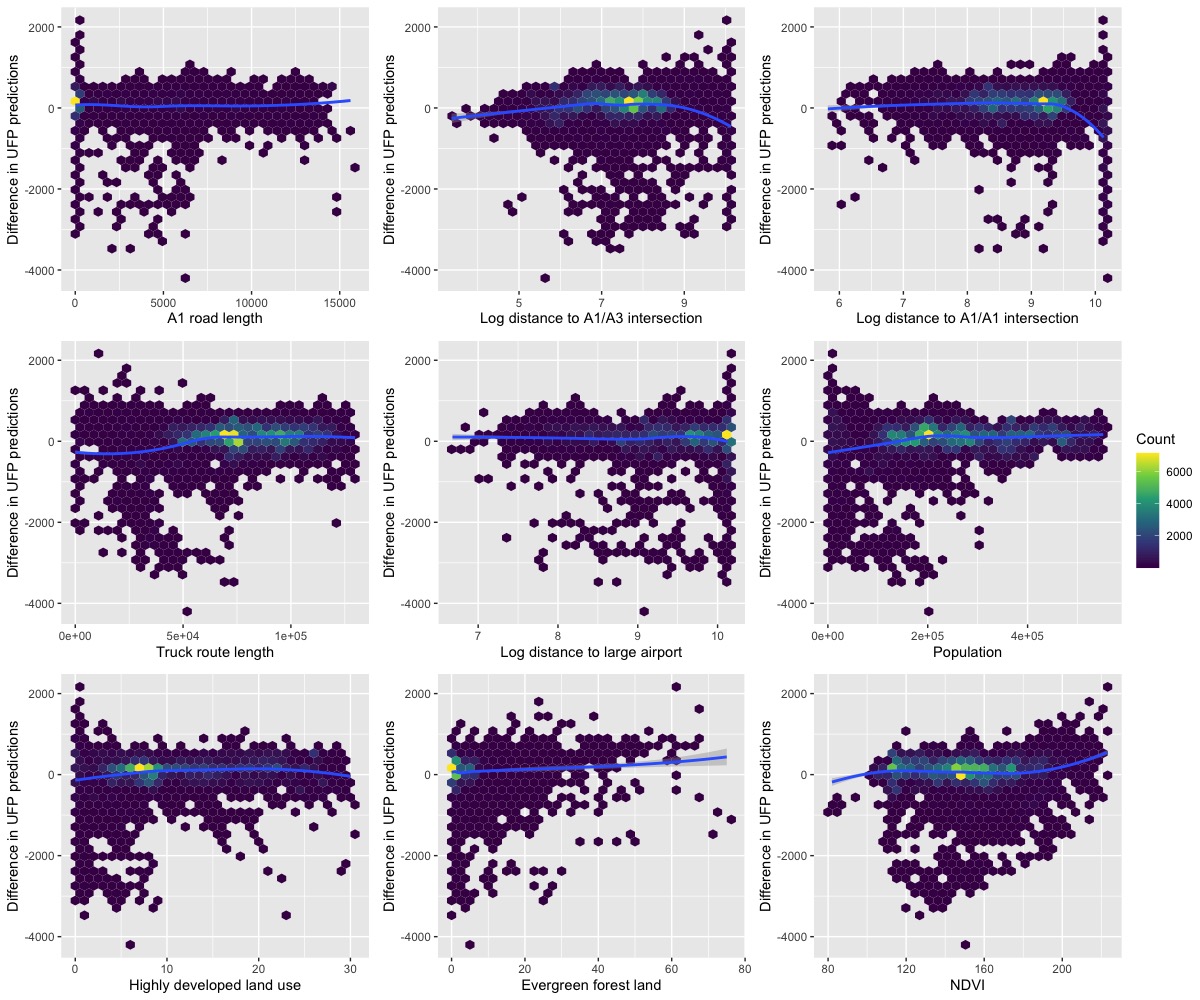}
    \caption{Hexagonal bin plot showing the difference between spatial RF (PL) and UK-PLS (the subtrahend) predictions of UFP concentration at the residential locations of an epidemiological cohort, versus the distribution of predictors with the greatest difference in variable importance between models. The color reflects the number of points falling to each small region of the plot. Locally weighted scatterplot smoothing (LOESS) curves are added to show the overall trend.}
    \label{fig:cohortufp_hexbin}
\end{figure}

\end{document}